\documentclass[%
 aip,
jap,
amsmath,amssymb,
groupedaddress, 
reprint,%
]{revtex4-1}

\usepackage{graphicx}
\usepackage{array}
\usepackage{dcolumn}
\usepackage{bm}
\usepackage[mathlines]{lineno}

\usepackage[utf8]{inputenc}
\usepackage[T1]{fontenc}
\usepackage{mathptmx}

\begin{document}

\title{A diamond anvil microassembly for Joule heating and electrical measurements up to 150 GPa and 4000 K}

\author{Zachary M. Geballe}
\author{Suzy M. Vitale}
\author{Jing Yang}
\author{Francesca Miozzi}
\author{Vasilije V. Dobrosavljevic}
\author{Michael J. Walter}
\affiliation{Earth and Planets Laboratory, Carnegie Institution for Science, Washington, DC 20015 
}%

\date{\today}

\begin{abstract}
When diamond anvil cell (DAC) sample chambers are outfitted with both thermal insulation and electrodes, two cutting-edge experimental methods are enabled: Joule heating with spectroradiometric temperature measurement, and electrical resistance measurements of samples heated to thousands of kelvin. The accuracy of temperature and resistance measurements, however, often suffers from poor control of the shape and location of the sample, electrodes, and thermal insulation. Here, we present a recipe for the reproducible and precise fabrication of DAC sample, electrodes, and thermal insulation using a three-layer microassembly. The microassembly contains two potassium chloride thermal insulation layers, four electrical leads, a sample, and a buttressing layer made of polycrystalline alumina. The sample, innermost electrodes, and buttress layer are fabricated by focused-ion-beam milling. Three iron samples are presented as proof of concept. Each is successfully compressed and pulsed Joule heated while maintaining a four-point probe configuration. The highest pressure-temperature condition achieved is~$\sim 150$~GPa and~$\sim 4000$~K. 
\end{abstract}

\maketitle

\section{Introduction}
Many breakthrough experiments have been enabled by innovative diamond anvil cell (DAC) loading techniques that combine two to four electrodes and one to two layers of thermal insulation. Two electrodes have been used for Joule heating (also known as ``internal resistive heating'') in order to study the melting curve of iron up to 25 GPa in one of the earliest DAC melting studies,\cite{Liu1975} and up to 270 GPa in one of the highest-pressure DAC melting studies.\cite{Sinmyo2019} Joule heated DACs have also been used to precisely map the hcp-fcc phase boundary of iron,\cite{Komabayashi2009} and to study the high pressure melting curves and/or equations of state of gold, tin, rhenium, and platinum.\cite{Weir2009, Weir2012, Zha2003, Zha2008, Geballe2021} Meanwhile, four electrodes have been used in electrical resistance measurements of iron and iron alloys compressed and laser-heated to 100s of GPa and 1000s of K in DACs in order to study the conductivity of Earth's core, with major implications for the history of Earth's geodynamo.\cite{Ohta2016, Zhang2020,Zhang2021,Zhang2022} Four-electrode loadings with thermal insulation have also been used to synthesize and characterize the electrical properties of new superconducting materials using laser-heating at 10s to 100s of GPa.\cite{Somayazulu2019,Drozdov2019,Zhu2023} Despite the decades-long publication record of Joule heated diamond cells and of electrical resistance measurements of thermally insulated diamond cell samples, the methods remain relatively uncommon and inaccessible because of the extreme difficulty of sample preparation.  

The main preparation challenge is to position the sample, electrical leads, and thermal insulation at the appropriate locations and with the appropriate orientations above the culet. First, two or four electrodes must connect to the sample of interest. During compression, the electrodes cannot translate too far from the edge of the sample, or cause too much deformation of the sample. For example, if an electrode slips inward, the shape of the sample of interest (e.g. the Joule-heated hotspot) can easily become too small or too irregular for accurate characterization. Second, the sample must be insulated from the diamond anvils with relatively uniform layers of a transparent, non-reactive insulator in order to limit temperature gradients as much as possible during Joule-heating or laser-heating. Otherwise, it is difficult to interpret spectroradiometric temperature measurements. For the case of four-point probe measurements, this typically means that six pieces of insulation - four electrical and two thermal - must be selected and placed in appropriate positions so that when pressure is applied, all insulation and electrodes remain well-positioned.

A set of publications from one to two decades ago presents an engineered solution for many applications requiring thermal insulation and electrodes. \cite{Weir2000,Velisavljevic2004,Weir2009,Weir2011,Weir2012,Lin2007} The solution is to synthesize ``designer diamond anvils'' using sputtered thin film electrodes, chemical vapor deposition (CVD) of an electrically insulating diamond layer, and polishing and etching to reveal the electrodes and create a pit for thermal insulation.\cite{Weir2009} Unfortunately, designer diamond anvils are not readily available (e.g. for purchase commercially) despite their invention more than ten years ago. Moreover, the electrical lead thickness reported for designer diamond anvils is less than  1~$\mu$m, limiting the amount of current that can be delivered.\cite{Weir2009}

Rather than using designer anvils, the thermally-insulated electrical experiments in Refs. \onlinecite{Liu1975, Boehler1986, Zha2008,  Komabayashi2009, Sinmyo2019, Geballe2021, Ohta2016, Ohta2010, Inoue2020, Zhang2020, Zhang2021, Zhang2022, Zhu2023} used standard diamond anvils and metal gaskets with electrically insulating inserts (or electrically insulating coatings). The typical description of the loading method in these studies lacks detail. For example, Ref. \onlinecite{Sinmyo2019} simply reports: ``The foil was loaded between the Al$_2$O$_3$ thermal insulation layers and connected to an electrode several millimeters away from the culet.'' Our previous publication using Joule-heated platinum samples provides slightly more detail, but it still lacks any prescription of how to appropriately position sample, electrodes, and insulation:\cite{Geballe2021} ``At least five pieces of platinum and several pieces of KCl are stacked so that when the diamond cell is closed, a central piece of platinum of 5 to 30 $\mu$m-width is separated from both anvils by KCl layers and electrically connected to the four outer electrodes.''  We suspect that the reason for the brief descriptions in many publications is that the actual methods rely on intuition, subtle choices of where to position materials, and lots of trial-and-error. 

Recently, Ref. \onlinecite{Ohta2023} presented an engineering solution to one technical problem by encapsulating a sample that is milled with a focused ion beam (FIB) in an insulator that is also milled with a FIB. Here, we also use FIB encapsulation of iron samples in four-point probe geometries. Moreover, we extend the encapsulation method in four ways: by using a three layer assembly method, presenting a standard recipe for fabrication of outer electrodes, employing Joule heating, and extending the pressure by using smaller culets. In addition, we document the reproducibility of our three-layer assembly method by reporting photos during and after heating of four samples -- three of which maintain a four-point probe geometry during compression and heating to~$> 2000$~K. No photo is saturated -- not even photos of hotspots that reach~$\sim 3000$~K peak temperature.

\section{Methods}
The method of sample preparation is divided into two sections: the fabrication of non-standard parts (\ref{section:fabrication}), and the assembly of all parts (\ref{section:assembly}). The two sections are presented in the chronological order (fabrication, then assembly), but they can be read in either order.  
The fabrication section describes recipes for four non-standard parts: an electrode holder and three thin slabs -- a bottom slab for thermal insulation, a middle slab that contains the sample, and a top slab for thermal insulation and electrical connection. The assembly section describes the procedure used to (a) make a gasket with an insulating insert, (b) integrate the gasket with outer electrodes, (c) stack the three thin slabs, (d) connect the top slab's electrode's to the outer electrodes, and (e) dehydrate and close the cell. 

The slabs and gasket use a homemade mixture of cubic boron nitride (cBN) and epoxy -- see Appendix \ref{section:cBN}. Hereafter, we refer to the mixture as ``cBN''.

All the ingredients (i.e. consumable materials) needed for the sample preparation method are listed in Table \ref{table:ingredients}. Finally, subsection \ref{section:compression_and_heating} briefly describes compression and heating of the sample.


\begin{table*}[ht]
\begin{tabular}{    p{4in}    c      c       }   
                                
\hline
Culet diameter ($\mu$m)  &  200   & 100    \\
Bevel diameter ($\mu$m) & NA & 300 \\    
Gasket thickness ($\mu$m)  &   28-34   &   20-24   \\
Side length of square gasket hole$^*$ &    140 &   90  \\
Side length of each slab$^*$ ($\mu$m) &   140 &   90  \\
Thickness of each slab ($\mu$m) &    8-11    &    6-8   \\
Hole diameter for KCl insulation$^*$ ($\mu$m) & 50 & 22 \\
Hole diameter for alumina insulation$^*$ ($\mu$m) & 70 & 30 \\
Cartesian coordinates of center of electrodes ($\mu$m) & ($\pm 40$,~$\pm 40$) & ($\pm 24$,~$\pm 24$) \\
Side length of square holes for electrodes$^*$ ($\mu$m) & 30 & 12 \\
\hline                              
\end{tabular}                                                               
\caption{Dimensions used for samples \#1, 3, and 4. Asterisks indicate nominal distances input into the laser mill. Our laser mill typically removes $\sim 2$~$\mu$m extra on each side our cBN pieces, and creates an edge with roughness~$\pm 2$~$\mu$m.}
\label{table:dimensions}
\end{table*}

\begin{figure*}[tbhp]
    \centering
    \includegraphics[width=.9\linewidth]{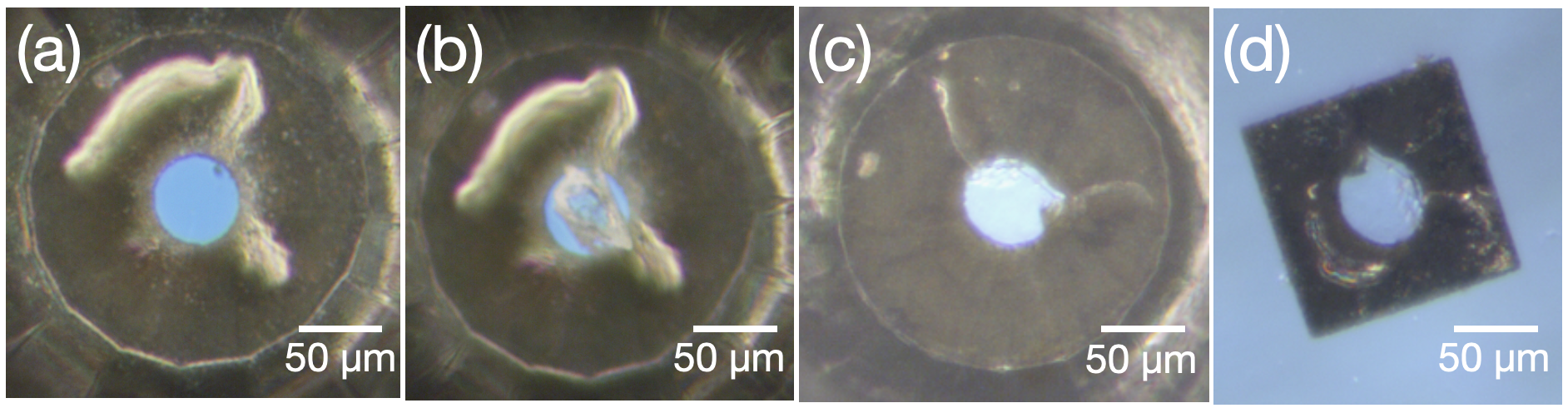}
    \caption{Fabrication of the bottom slab for sample \#1. (a) A disc of cBN pressed into a gasket, with a~$\sim 50$~$\mu$m diameter hole. The hole was laser milled after indentation to 8-11~$\mu$m thickness. (b) After a chunk of KCl is placed in the hole. (c) After pressing the KCl to fill the hole. (d) After milling out the central square and transferring it to a glass slide.}
    \label{fig:bottom_slab}
\end{figure*}

\begin{figure*}[tbhp]
    \centering
    \includegraphics[width=.9\linewidth]{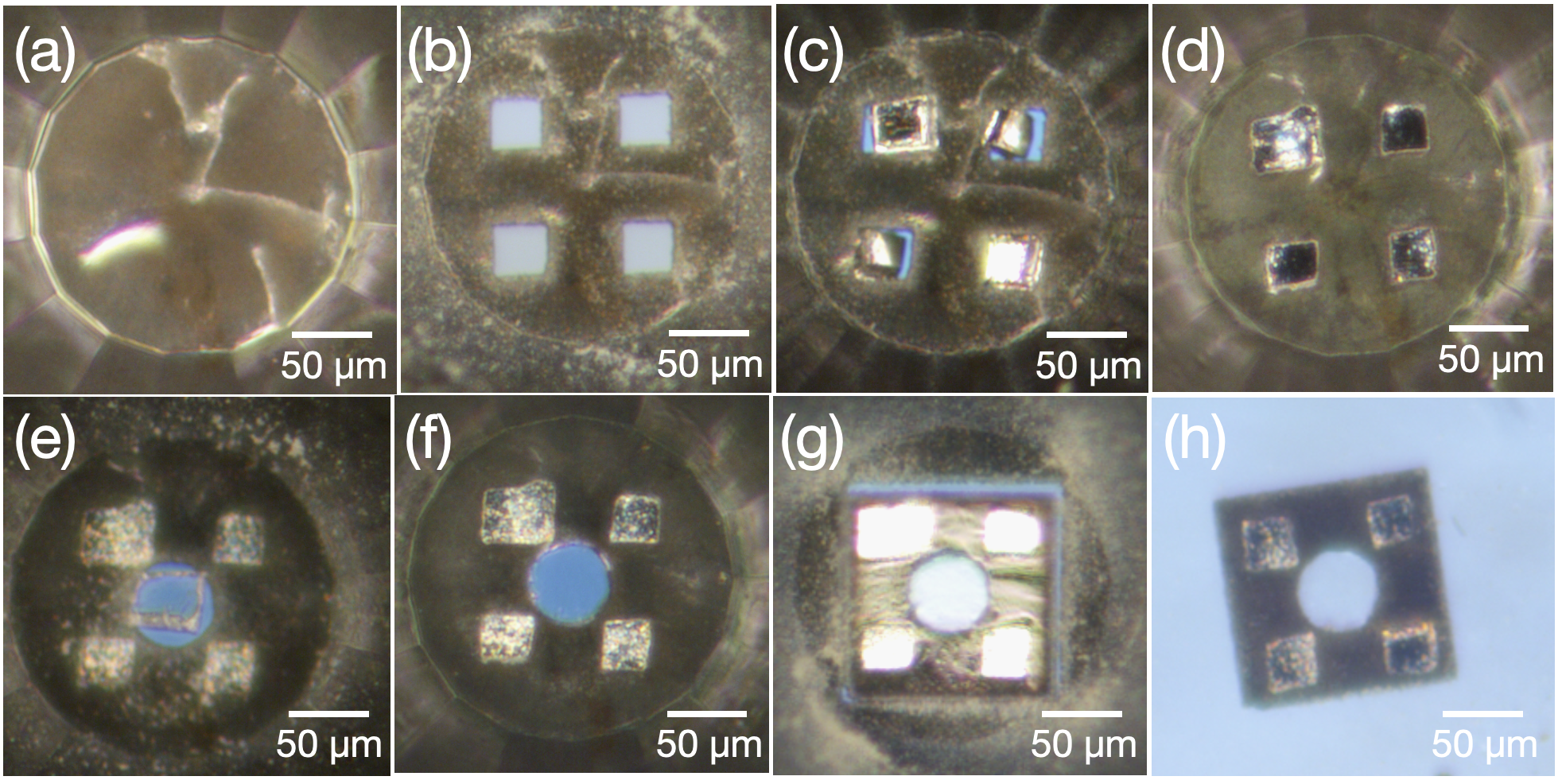}
    \caption{Fabrication of the top slab for sample \#1. (a) An 8-11~$\mu$m-thick cBN disc pressed into a gasket. (b) After milling four holes: $30 \times 30$~$\mu$m squares located at Cartesian coordinates ($\pm 40$~$\mu$m$^2$, $\pm 40$~$\mu$m$^2$). (c) After resting four platinum squares on top of the holes. (d) After pressing the platinum into the holes with a DAC. (e) After milling a 60~$\mu$m-diameter hole and resting a square-shaped piece of KCl in the hole. (f) After pressing the KCl into the hole. (g) After milling a~$140 \times 140$~$\mu$m$^2$~square centered around the KCl. (h) After moving the central square to a glass slide, and accidentally flipping upside down.}
    \label{fig:top_slab}
\end{figure*}

\begin{figure*}[tbhp]
    \centering
    \includegraphics[width=.9\linewidth]{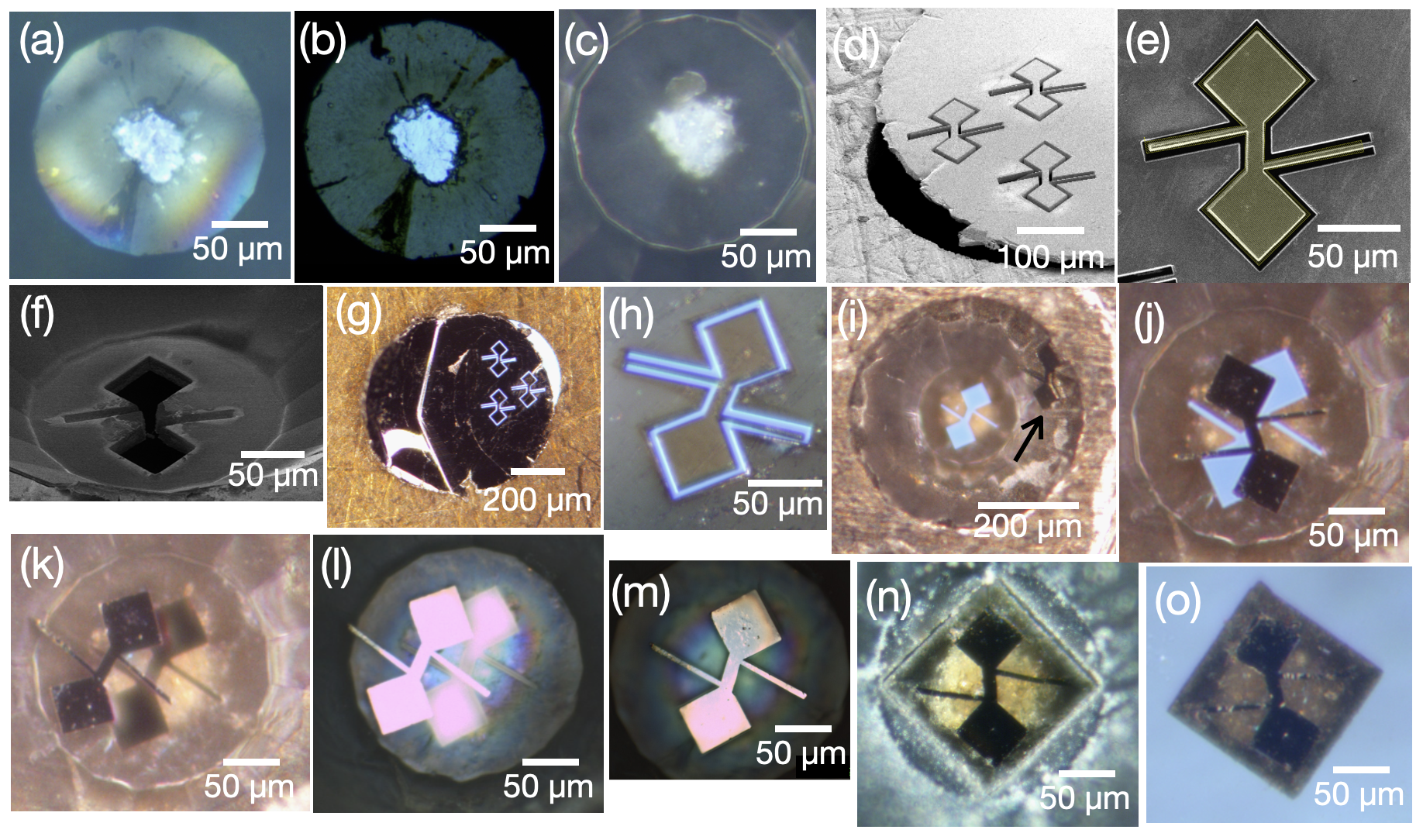}
    \caption{Fabrication of the middle slab for sample \#1. (a-b) Photos during compression to~$\sim 30$~GPa of an 8-11~$\mu$m-thick cBN insert with a $\sim 70$~$\mu$m region of alumina powder, viewed in reflected light (a) and in transmitted light (b). (c) After decompression to ambient pressure. (d) An SEM image of three iron samples. (e) An ion beam image of one iron sample overlaid with a yellow FIB patterning region that is at least 1~$\mu$m wider than the sample in all regions. (f) The cBN and alumina from (c) after FIB milling of the yellow pattern region from (e). (g-h) Optical images of the iron samples. (i) After removing one iron sample by laser milling the attached arm, and transferring it to a location near the cBN and alumina (black arrow). (j) After micromanipulating to place the sample above the hole. (k) After placing a bar magnet below the anvil to orient the iron sample in line with the hole. (l) Image from the micromanipulator. (m) After micromanipulating to gently press the sample into the hole. (n) After pressing the sample fully into the hole with a DAC and milling a~$140 \times 140$~$\mu$m$^2$~square around the sample. (o) After moving the central square to a glass slide.}
    \label{fig:middle_slab}
\end{figure*}

\begin{figure*}[tbhp]
    \centering
    \includegraphics[width=\linewidth]{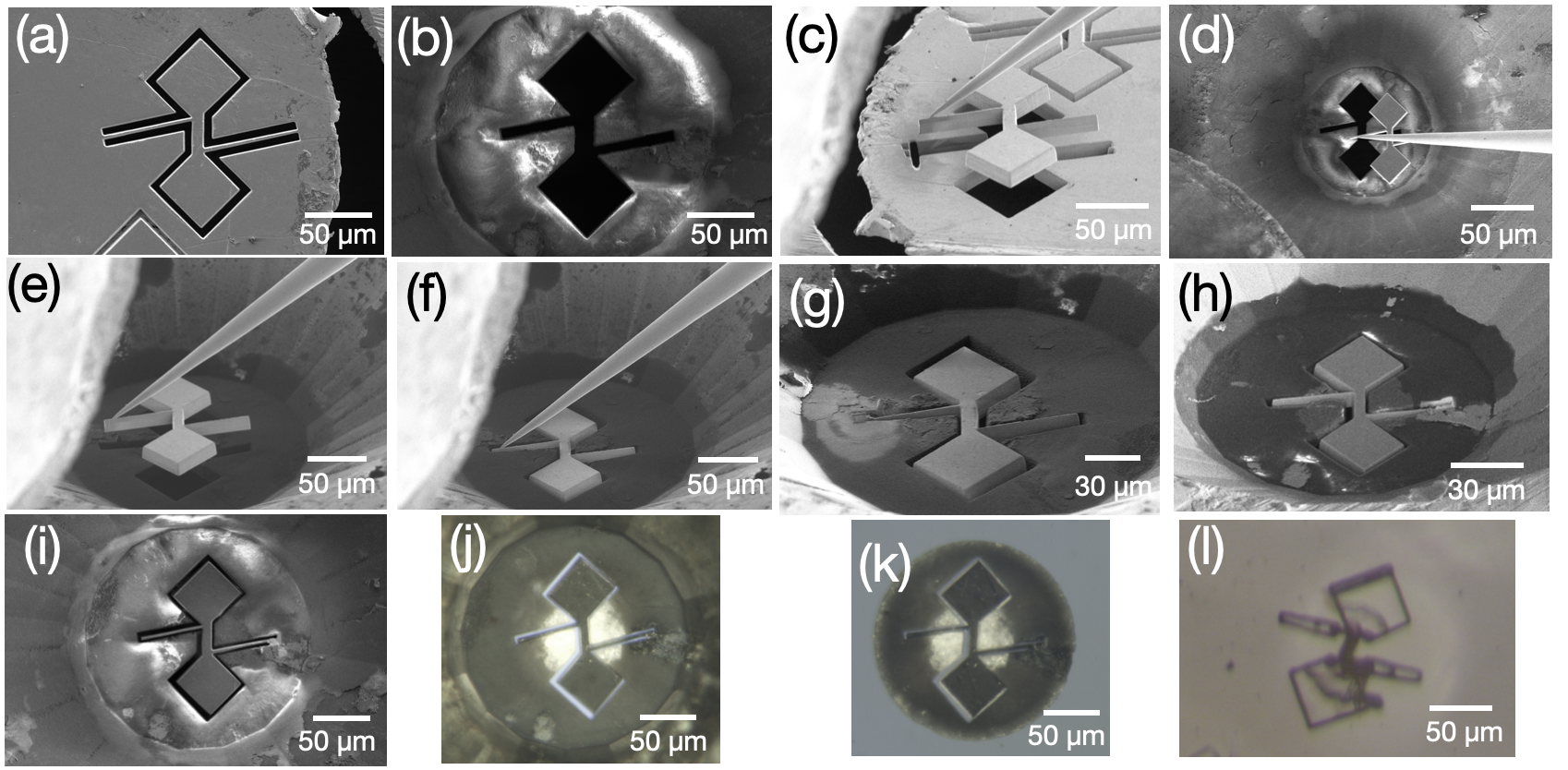}
    \caption{Fabrication of the middle slab using the FIB ``lift-in'' procedure for sample \#2. Here, the final slab is a circle, rather than a square. (a) An SEM image of an iron sample, collected using the SEM column in the dual beam FIB/SEM instrument. (b) An SEM image of a hole in the cBN/alumina slab (see Fig. \ref{fig:middle_slab} for fabrication details). (c) An ion beam image after using tungsten deposition to weld the sample to the tungsten needle that is held by the FIB's micromanipulator arm, cutting the iron sample free, and lifting the sample by~$\sim 20$~$\mu$m. (d) An SEM image of the sample and needle near the hole in the cBN. (e) An ion beam image after moving the sample very close to the hole. (f) After moving the sample into the hole. (g) After using tungsten deposition to weld one thin arm of the sample to the hole, milling the needle free, and retracting the needle. I.e., the reversal of the process in (c). (h) An SEM image after rastering the ion beam across the surface to clean off stray tungsten deposition. (i) SEM top view. (j) Optical image after removing from the FIB. (k) After laser milling to remove the disc. (l) A single crystalline slab of alumina used as a backing plate to avoid contamination of the cBN and alumina piece during FIB milling, viewed after milling.}
    \label{fig:FIB}
\end{figure*}

\subsection{Fabrication of non-standard parts}
\label{section:fabrication}

\subsubsection{Fabrication of the bottom, middle, and top slabs}
\label{section:slabs}

The fabrication and use of three thin slabs are the main innovation in this paper. The three slabs contain the sample, thermal insulation, electrical insulation, the innermost parts of the gasket, and the innermost electrodes.

Slab fabrication is accomplished using an auxiliary diamond cell with 200~$\mu$m~culets and three spare steel gaskets. Our auxiliary diamond cell is a smooth-sliding ``symmetric cell''. The key requirements for the auxiliary cell is that it (a) opens and closes easily, and (b) reproducibly compresses samples to at least 30 GPa without any need for realignment of anvils.

The first step in fabricating each slab is to press a cBN disc inside a steel gasket. First, a steel gasket is compressed to 20 GPa, and the entire culet region (200~$\mu$m-diameter) is removed by laser milling. Next, a chunk of cBN is pressed into the steel gasket. The cBN is thinned by an iterative process of (a) laser milling a hole (e.g.  120~$\mu$m-diameter) and (b) compressing to 10-30 GPa. The target thickness is slightly less than 1/3 the thickness of the indentation in the real gasket that will be used in the high pressure experiment. For example, three 10~$\mu$m slabs are ideal for a 32~$\mu$m-thick indentation used for experiments on 200 $\mu$m culets. Alternatively, some slabs can be be slightly thicker and others slighter thinner, as long as the sum of the three slab thicknesses is slightly less than the thickness of the gasket for the actual experiment. Table \ref{table:dimensions} lists dimensions for gaskets and slabs. To measure thickness accurately, we either (a) recover a laser-milled disc of cBN, turn it on its side, and measure with a microscope, or (b) gently press the gasket between anvils and measure with interferometry using white light and a spectrometer. To speed up the indenting process, we use a torque measuring screwdriver to apply consistent torque (e.g. 4 inch-pounds) on each screw.

The bottom slab is used for thermal insulation. It contains a cBN outer region and a KCl-filled hole. Photos of the bottom slab for sample \#1, the sample that is highlighted here, are shown in Fig. \ref{fig:bottom_slab}. After preparing a 10~$\mu$m-thick, 200~$\mu$-diameter cBN slab at the center of a steel gasket, we use a laser mill to drill a 50~$\mu$m diameter hole, place a small piece of KCl in the hole, close the DAC, and press down by hand (i.e. without using screws for compression). More KCl can be added if the original piece fills less than $\sim 80$\% of the hole. Overfilling the hole can lead to a gasket blowout during the compression experiment. Note that our laser mill uses a sub-nanosecond pulsed near-IR laser (PowerChip PNP-M08010) and follows the design of Ref. \onlinecite{Hrubiak2015}.

The top slab is used for thermal insulation and electrodes. The process is similar to the process for the bottom slab, but with extra steps for electrode placement (Fig. \ref{fig:top_slab}). First, four squares are laser milled at the Cartesian coordinates ($\pm 40$~$\mu$m, $\pm 40$~$\mu$m) with respect to the slab center. Each square has a 30~$\mu$m side-length. Next,~$\sim 10$~$\mu$m-thick platinum is cut into~$\sim 30 \times 30$~$\mu$m$^2$ squares and placed in the holes. (All pieces of platinum in this paper are cut by hand - we use straight razors for all pieces $<100$~$\mu$m diameter and scissors for 127~$\mu$m diameter wire.)

The middle slab contains the sample, innermost electrodes, and the sample's lateral buttress. In this study we use iron as the sample, and fabricate it along with the innermost electrodes from a single slab or iron. The buttress material is alumina for all examples in the main text; in appendix \ref{section:pure_media}, one example uses KCl. 

Fig. \ref{fig:middle_slab} shows the middle slab fabrication process for sample \#1. The first steps are similar to fabrication of the bottom slab: a 10~$\mu$m-thick cBN slab is created, a 70~$\mu$m diameter hole is milled in the cBN, and alumina powder (Johnson Matthey 22 $\mu$m alpha, 99.5\% metals basis) is pressed into the hole. Unlike in the case of KCl insulation, alumina poses little risk of gasket blowout because of its high yield strength. Still, the relatively brittle alumina region should be small enough to allow easy maneuvering of the final~$140 \times 140$~$\mu$m$^2$~slab without crumbling. Next, sample material is prepared for FIB milling. The iron starting material is pressed to 10~$\mu$m thickness using a second auxiliary DAC outfitted with 1 mm culets. The 10~$\mu$-thick piece of iron is then glued over a hole in a 400 $\mu$m-thick brass slab using a~$\sim 200$~$\mu$m blob of silver epoxy (or any other sticky material that is somewhat electrically conductive). Next, the brass slab is clamped atop a~$\sim 5 \times 5$~mm$^2$ piece of iron foil on an aluminum SEM pin stub. Likewise, the steel gasket with cBN/alumina slab is clamped atop a piece of sapphire on an aluminum SEM pin stub. The use of the~$5 \times 5$~mm$^2$~iron foil and sapphire piece at the bottom of the sample are crucial for preventing contamination of the iron and sapphire during FIB milling.

Next, a FIB (Helios G4 PFIB UXe DualBeam FIB/SEM) is used to shape the iron slab into a sample and innermost electrodes in a four-point probe configuration, and to mill a matching hole in the cBN/alumina slab. Using a 4 nA ion beam, 8~$\mu$m-wide trenches are cut through the iron to make the shape in Figs. \ref{fig:middle_slab}d-e. The shape contains a central sample that is~$8 \times 22$~$\mu$m in surface area, along with four electrodes. The resulting shape of iron is recorded in an SEM image collected using the SEM column of the dual beam instrument. Note that a connection is maintained between the iron piece and the main iron foil during this step of milling (Fig. \ref{fig:middle_slab}d). A slightly wider shape (e.g.~$0.5$~$\mu$m wider in each dimension) is used as the pattern for milling the hole in the cBN/alumina slab (yellow pattern in Fig. \ref{fig:middle_slab}e). The shape is milled with 4 nA or 15 nA ion beams, depending on sample dimensions and time constraints. Fig \ref{fig:middle_slab}f,i,j shows the results for sample \#1. Fig. \ref{fig:FIB}b show the result for sample \#2.

Finally, the iron is transferred into the alumina hole using one of two methods, shown in Fig. \ref{fig:middle_slab} and Fig. \ref{fig:FIB}, respectively. In the case of samples \#1, 3, and 4,  the transfer is performed outside the FIB (e.g. Fig. \ref{fig:middle_slab}i-m). First, we seat the steel gasket with cBN/alumina onto the auxiliary diamond cell. During this step, we do not close the auxiliary cell; doing so would deform the hole in the cBN and alumina. Next, we use the laser mill to cut the connection with the main platelet and free the Fe sample . We then transfer the iron to the cBN/alumina slab by hand (Fig. \ref{fig:middle_slab}i), and use a micromanipulator (Microsupport AxisPro APSS) to push the sample on top of the hole (Fig. \ref{fig:middle_slab}j). For some samples, including sample \#1, alignment of the sample and the matching hole is a challenge because of magnetic interaction between the iron sample on the one hand, and the tungsten carbide seat and/or the diamond cell body on the other hand. In such cases, we remove the DAC from the micromanipulator, place a neodymium magnet below the anvil, and orient the magnetic field along the long direction of the hole in the cBN/alumina slab. As long as the sample does not fly away during this process, the result is a well-oriented sample (Fig. \ref{fig:middle_slab}). Next, we return the DAC to the micromanipulator (Fig. \ref{fig:middle_slab}l), move the sample so that it is resting atop of the hole, and press gently downwards on the large areas of the iron so that the sample is stuck partway into the hole (Fig. \ref{fig:middle_slab}m). Now that the sample is well positioned, we close the cell. As it closes, a relatively uniform force is applied across the iron sample, pressing the iron it into the hole in the cBN/alumina. As we press with more force, the cBN, alumina, and/or iron deform slightly to eliminate air-gaps. Finally, the middle slab is cut free from the auxiliary steel gasket using the laser mill (Fig. \ref{fig:middle_slab}n-o). 

In the case of samples \#2, 5, 6, and 7, the FIB's in-situ micromanipulator is used to place the iron in the hole (\ref{fig:FIB}) in a procedure we call a ``lift-in'', in analogy with the well-known ``lift-out'' FIB procedure for extracting thin sections. For sample \#2, we used circular slabs (i.e. discs) rather than square slabs. The ``lift-in'' recipe is: (1) The FIB's micromanipulator is outfitted with a sharp tungsten needle. The needle's tip is positioned just above the electrode that is still attached to the bulk of the iron slab. (2) The needle's tip is attached to the iron electrode using a small ($\sim 1 x 5$~$\mu$m$^2$)~pad of tungsten deposition (WCO). (3) The iron is cut free using 15 nA beam current, and lifted vertically away from the bulk of the iron slab (Fig. \ref{fig:FIB}d). (4) The needle is retracted. (5) The stage is moved to the cBN/alumina position. (6) The needle with iron sample is re-inserted and lowered into the hole in the cBN/alumina slab(Fig. \ref{fig:FIB}e-g). (7) Tungsten deposition is used to attach the iron to the cBN. (8) The needle is milled free, and retracted. An SEM image shows that the tungsten deposition extends in a $\sim 50 \times 80$~$\mu$m oval, far beyond the intended weld rectangle ( Fig. \ref{fig:FIB}g). To minimize contamination, we typically raster an ion beam across the entire cBN/alumina slab area except for the weld point. The final product is shown in Figs. \ref{fig:FIB}h-k. 

After removing the sample from the FIB and from the aluminum stub holder, we confirm that the iron and alumina pieces that were underneath the iron foil and the alumina plus cBN layer were indeed milled by the ion beam (e.g. Fig. \ref{fig:FIB}l). Otherwise, a contamination layer of some other stub material (e.g. aluminum) is created on the underside of the iron sample or alumina buttress.

The two methods of transferring iron into the alumina hole have similar success rates in our experience. The methods require different skills -- skills using a FIB and its micromanipulator in one case; skills using needles by hand and a stand-alone micromanipulator in the other case.

\begin{figure*}[tbhp]
    \centering
    \includegraphics[width=5in]{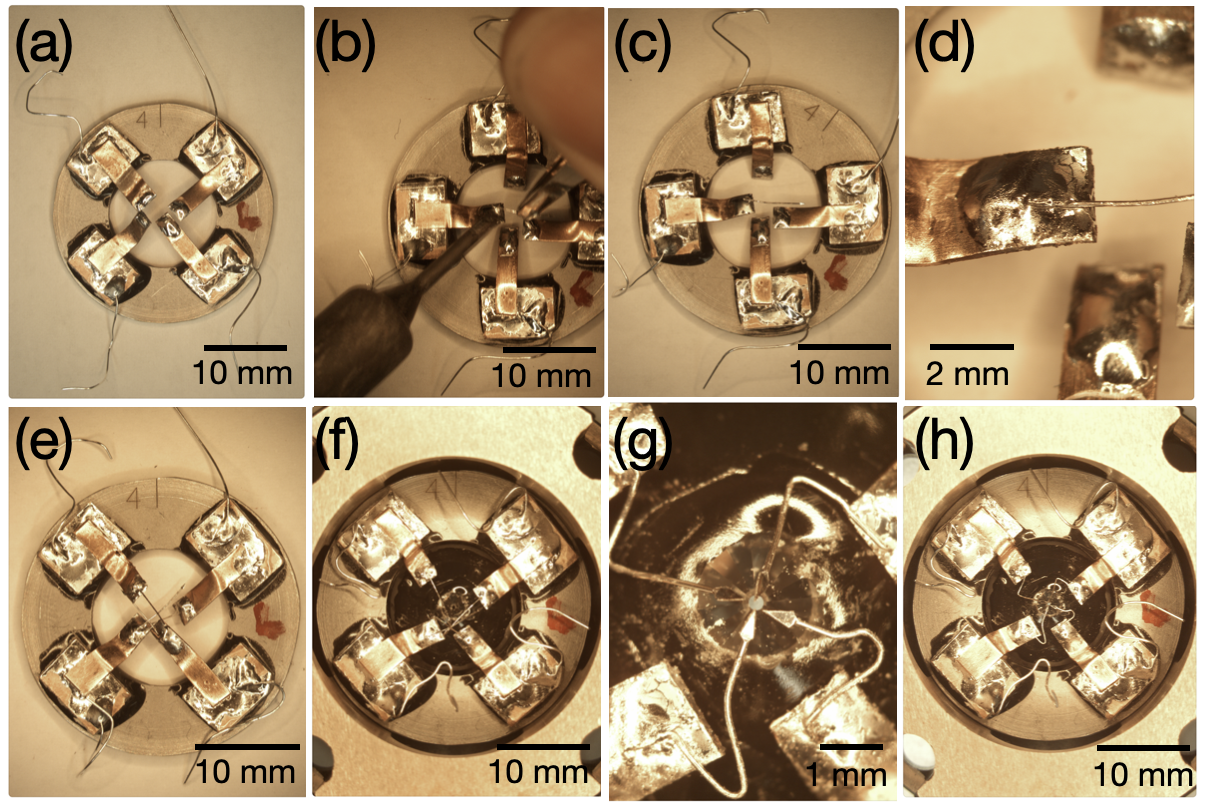}
    \caption{Fabrication of the electrode holder for sample \#2. (a) The 1.5 mm-thick stainless steel disc after using epoxy to glue four rectangles of copper-clad board and soldering four pieces of copper foil and four tinned copper wires. (b) An action photo of soldering the first platinum wire. (c-d) After soldering the first platinum wire to the free end of a piece of copper foil. (e) After soldering all four platinum wires. (f) After sliding the electrode holder into inside the DAC piston's cavity. (g-h) After bending the pieces of copper foil and platinum wire so that all four arrowheads are within~$\sim 100$~$\mu$m of the culet's edge.}
    \label{fig:electrode_holder}
\end{figure*}

\subsubsection{Fabrication of the electrode holder}
\label{section:electrode_holder}
An electrode holder that is independent of the gasket (i.e. never glued to the gasket) is a helpful tool for creating a robust electrical pathway from the edge of the cell to the edge of the culet. Our early versions of electrode holders are described briefly in Refs. \onlinecite{Somayazulu2019, Geballe2021}. Here, we fabricate the electrode holder using the following procedure (Fig. \ref{fig:electrode_holder}): (1) A 1.5 mm-thick stainless steel disc is made using standard machine shop techniques. The outer diameter is~$\sim 27$~mm; each disc is machined to match the inner diameter of the piston of the diamond cell. The disc's inner diameter is 13 mm, allowing plenty of clearance around the tungsten carbide seat. (2) Four sides of the disc's outer diameter are further machined using 240 grit sandpaper in order to add clearance when sliding the disc into the cell. The four flattened sides are designed to align with the portholes of the diamond cell. (3) Four rectangles of copper-clad board ($7 \times 5$~mm, 1/16 inch-thick) are glued to the disc using epoxy (Stycast 2750FT). The positions for the copper boards are chosen to avoid obscuring the view through the diamond cell's portholes. In addition, epoxy is added as electrical insulation along the inner diameter of the stainless steel ring. (4) Four copper wires (0.2 mm diameter, 2 cm long) are soldered to the copper board. (5) Four pieces of copper foil ($ 9 \times 2 \times 0.2$~mm) are soldered to the copper board, pointing inwards. The result is shown in Fig. \ref{fig:electrode_holder}a. (6) Four platinum wires are soldered to the copper foil using a very small amount of solder. Before soldering, it is best practice to shape each end of the platinum wire. The outer end is pressed slightly in order to ease the wetting of the platinum by solder. The inner end is pressed to~$\sim 40$~$\mu$m-thickness and razor-cut to an arrow shape that is $\sim 400$~$\mu$m-long and tapered from~$\sim 200$~to less than 50~$\mu$m width at its tip. The process and result are shown in Fig. \ref{fig:electrode_holder}b-f. Note that the order of steps (5) and (6) can be reversed. (7) The electrode holder is placed in the DAC, and secured with sticky tack, if necessary. (8) Tweezers are used to bend the copper foil and platinum wires until the tips of the arrowheads are within $\sim 120$~$\mu$m of the tip of the culet. The result is shown in Fig. \ref{fig:electrode_holder}g,h. 

Note that the platinum wire is purposely chosen to be almost double the length needed to reach the culet's edge. The extra length allows the platinum to be shaped into an S-curve which can be bent in order to adjust the position of the platinum arrow. The extra length also allows re-use of the outer electrodes  -- see Appendix \ref{section:re-use} for details.

\begin{figure*}[tbhp]
	\centering
	\includegraphics[width=.9\linewidth]{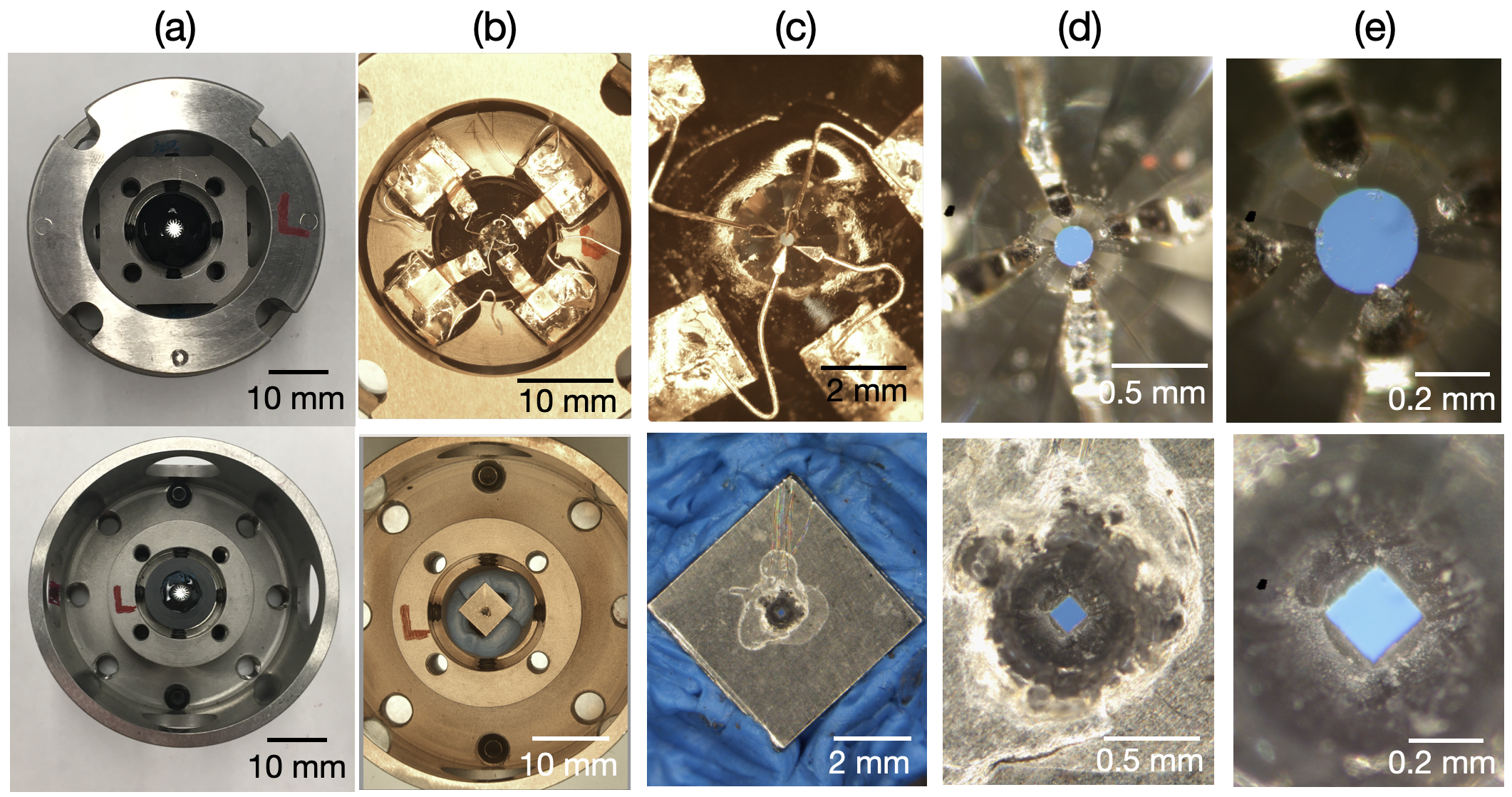}
	\caption{Representative photos of a DAC piston (top row) and cylinder (bottom row) ready to be outfitted with a microassembly. (a) A Zha-type DAC with tungsten carbide seats and diamond anvils. (b-c) After adding the outer electrode holder and a gasket with cBN insert. (d-e) Magnified view of near of sample \#1 with its gasket and well positioned outer electrodes.}
	\label{fig:outer_stuff}
\end{figure*}

\subsection{Assembly}
\label{section:assembly}
The assembly recipe assumes successful fabrication of the bottom, middle, and top slabs (Section \ref{section:slabs}; Figs. \ref{fig:bottom_slab}d, \ref{fig:middle_slab}o, and \ref{fig:top_slab}h), as well as the electrode holder (Fig. \ref{fig:electrode_holder}h). Photographs of the assembly process for sample \#1 are in Figs. \ref{fig:outer_stuff}, \ref{fig:stack}, and \ref{fig:closing}. Photographs of the fabricated slabs and the assembly process for sample \#4, which was loaded on 100~$\mu$m culets, are shown in Fig. \ref{fig:100um}.

We divide the assembly description into three parts: (a) assembly outside the culet region, (2) assembly on top of the culet, and (3) completing the circuit and closing the DAC.

\subsubsection{Assembly outside the culet region}
We use Zha-type DACs.\cite{Zha2017} Other types of cells can also be used -- see Appendix \ref{section:other_cells}. We use standard anvils (2 mm tall; standard cut; 200~$\mu$m flat culet or 100~$\mu$m flat with a single 8 degree bevels to 300~$\mu$m diameter), standard seats (tungsten carbide; 1 mm opening; 60 degree full opening), and a standard gasket material (250~$\mu$m-thick tungsten or rhenium). We use standard procedures to glue and align the anvils and to indent the tungsten gasket (20 and 30 GPa indentation pressure for 200 and 100~$\mu$m culets, respectively). The gasket is held with sticky tack onto the cylinder side of the DAC (Fig. \ref{fig:outer_stuff}).

An electrically insulating insert made of cBN is fabricated in the tungsten gasket by the following procedure. First, we laser mill a hole that covers the entire culet and bevel region (e.g. a 300~$\mu$m diameter hole for a 300~$\mu$m diameter bevelled area). Next, we add a large, $\sim 400$~$\mu$m diameter chunk of cBN, and compress to 30-35 GPa. Pressure is measured using ruby fluorescence or the diamond anvil Raman edge.\cite{Mao1986,Akahama2006} Additional cBN is added and the gasket is re-indented until three conditions are met: (a) cBN completely insulates the tungsten indentation from the piston-side anvil, (b) the thickness of cBN insulation at the top of the indentation is at least 40~$\mu$m, and (c) the cBN sticks to the tungsten gasket when the cell is opened (rather than sticking to the piston anvil). Typically, three to ten iterations of ``add cBN'' and ``indent to 30-35 GPa'' are required to achieve all conditions. After achieving all conditions, the cBN thickness can typically be reduced by further indentation without ruining any of the conditions (a)-(c). The indentation thickness is 28-34~$\mu$m for 200~$\mu$m culets and 20-24~$\mu$m for 100~$\mu$m culets. After fabricating the cBN insert, we add a ring of glue (Loctite Gel Control Super Glue) or epoxy (Stycast 2750FT) around the outermost cBN. For example, Fig. \ref{fig:outer_stuff}d shows a the ring of transparent glue encircling the cBN. The ring improves the mechanical stability of any pieces of cBN that are sticking up above the tungsten surface, as well as electrical insulation to separate platinum electrodes from the tungsten surface. 

An electrode holder with outer electrodes is created by the recipe described in section \ref{section:electrode_holder}, and placed into the inner diameter of the DAC's piston (Fig. \ref{fig:outer_stuff}b). The steel disc rests flatly against the platform that houses the set screws for the tungsten carbide seat. To secure the disc, we press centimeter-sized pieces of sticky tack against the edges of disc and against the piston. The center of the electrode holder contains four platinum wires with tips that have been shaped into arrowheads and positioning withing~$\sim 120$~$\mu$m of the edge of the 200~$\mu$m diameter culet, as shown in Fig. \ref{fig:outer_stuff}c. 

Next, we integrate the outer electrodes and the gasket, using an iterative ``press and bend'' process. We press the platinum into the cBN, rearrange the platinum arrowheads by using tweezers to bend the platinum and the copper foil that is soldered to the platinum, and iterate many times. Once the four platinum arrowheads are well positioned (within~$\sim 120$~$\mu$m of the culet edge, and well separated from each other) and do not move substantially when the DAC is closed, the outer electrode ``press and bend'' operation is complete (Fig. \ref{fig:outer_stuff} d-e).

Next, we cut a square-shaped hole at the center of cBN insert. The side length of the square is 140~$\mu$m for the 200~$\mu$m culets and 90~$\mu$m for the 100~$\mu$m culets. We use a square hole instead of a circle to simplify rotational alignment of the pieces that are placed inside the hole (see section \ref{section:slabs}) -- an especially useful trick for for the 100~$\mu$m culets.

At some time before closing the DAC for the final time, the four copper wires that are attached to the edge of the electrode holder are soldered to four pieces of copper-clad board (~$5 \times 5$~mm, 1/32 inch-thick) that are glued with epoxy to the cylinder's portholes. It is important to solder while the DAC is closed so that vapors from soldering flux do not precipitate onto the anvil's culet. 

At this point, all preparations are complete for the region that is outside the culet region and inside the DAC piston and cylinder. The result is a gasket, cBN gasket insert, and four isolated paths of metal (copper, solder, and platinum) connecting four points on the edge of the Zha-cell piston to four points near the edge of the culet.

\subsubsection{Assembly on top of the culet}

\begin{figure*}[tbhp]
    \centering
    \includegraphics[width=.95\linewidth]{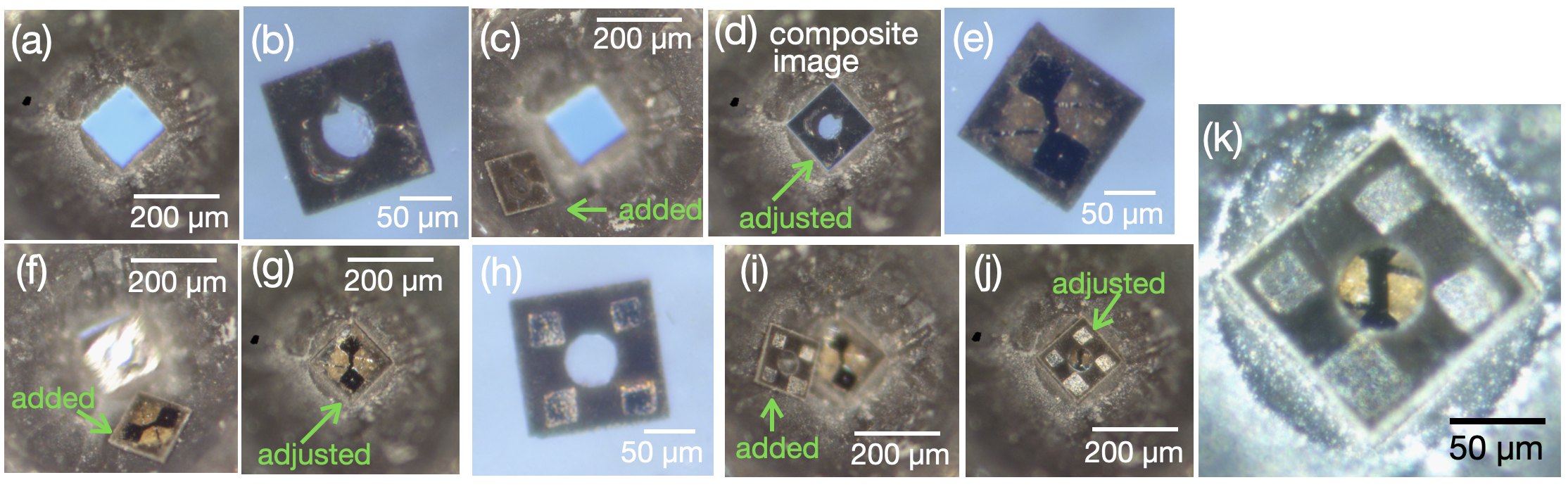}
    \caption{Stacking the bottom (b-d), middle (e-g), and top slabs (h-j). Green annotations indicate completed processes  --  a slab that was ``added'' by hand, or a piece that was ``adjusted'' using a micromanipulator. Note that (d) is a composite image made from images (a) and (b); all other images are real photos. (k) The final frame is a high-magnification photo that shows the central portion of the middle slab viewed through the transparent KCl region of the top slab.}
    \label{fig:stack}
\end{figure*}

One by one, we transfer each of the three thin slabs onto the cBN gasket insert and using a micromanipulator, push each delicate slab into the hole of the gasket (Fig. \ref{fig:stack}). After stacking the top slab, we close the cell and press firmly by hand to make sure the stack of slabs is seated inside the gasket hole.

\subsubsection{Completing the circuit and closing the DAC}
\label{section:completing_the_circuit}

\begin{figure*}[tbhp]
    \centering
    \includegraphics[width=\linewidth]{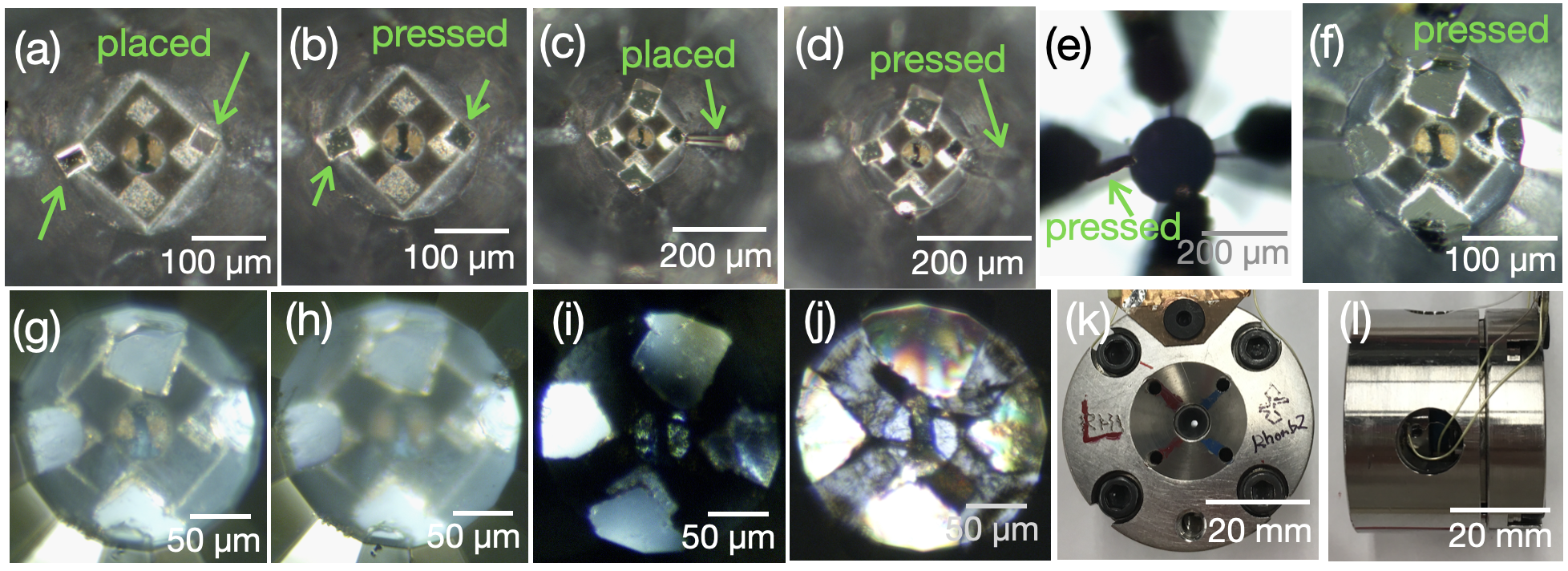}
    \caption{Completing the circuit and compressing the DAC. (a-f) Completing the circuit by placing and pressing~$\sim 10$~$\mu$m-thick pieces of platinum. Green annotations mark pieces of platinum that have been ``placed'' or ``pressed'' between the anvils. Panel (e) shows the cylinder side anvil with four wide electrodes on the top, bottom, left, and right, and one narrow electrode indicated by the green arrow; all other panels show the piston side. (g-i) Closing the DAC and compressing to 4 GPa. Panels (g) and (h) use two different focal planes, showing that the cell is slightly open and ready to be dehydrated in a vacuum oven. (j) After compressing to 50 GPa. (k-l) The DAC after fastening a copper-clad board to the outside of the DAC and soldering four electrical connections. }
    \label{fig:closing}
\end{figure*}

Next, we complete the four-point probe circuit, and close the DAC. To complete the circuit, the four small squares of platinum shown in Fig.~\ref{fig:stack}k must be connected to the outer electrodes shown in Fig.~\ref{fig:outer_stuff}. We make the connection by a ``place and press'' iterative procedure. We place pieces of platinum, then press them into the cBN gasket and slabs by closing the DAC, and repeat many times (Fig.~\ref{fig:closing}). We use a combination of 25~$\mu$m diameter platinum wire (e.g. Fig.~\ref{fig:closing}c) and 5 to 10~$\mu$m-thick slabs of platinum (e.g. Fig.~\ref{fig:closing}a), which are themselves made by pressing the 25~$\mu$m wire to the desired thickness between 1 mm anvils in the second auxiliary DAC. The three keys to success in this step are: (1) ensure that pieces of platinum are never pressed against the KCl at the center of the culet, (2) continue placing and pressing small pieces of platinum until the four-point probe is complete and until the pieces do not shift out-of-position when gently pressed between the anvils, and (3) make sure that each electrical path is narrow enough to avoid short circuits during the compression experiment. 

Finally, the DAC is closed most of the way, leaving a $\sim 5$ to 20~$\mu$m gap for gas flow (Fig.~\ref{fig:closing}g-h). The DAC is inserted into a vacuum oven at 115$^\circ$~C for 1 hour to dehydrate the KCl. After 1 hour, the oven is purged with argon gas, and within~$\sim 5$~seconds, the cell is removed and each of two screws is tightened to 1.5 in-lbs of torque. After cooling to room temperature, the measured pressure is typically in the range 2-5 GPa (Fig.~\ref{fig:closing}i). In some cases, platinum connectors slip during cell closure, creating open circuits or short circuits. In these cases, the cell can typically be opened, platinum can be replaced or rearranged, and the dehydration procedure can be repeated.

\begin{figure*}[tbhp]
    \centering
    \includegraphics[width=\linewidth]{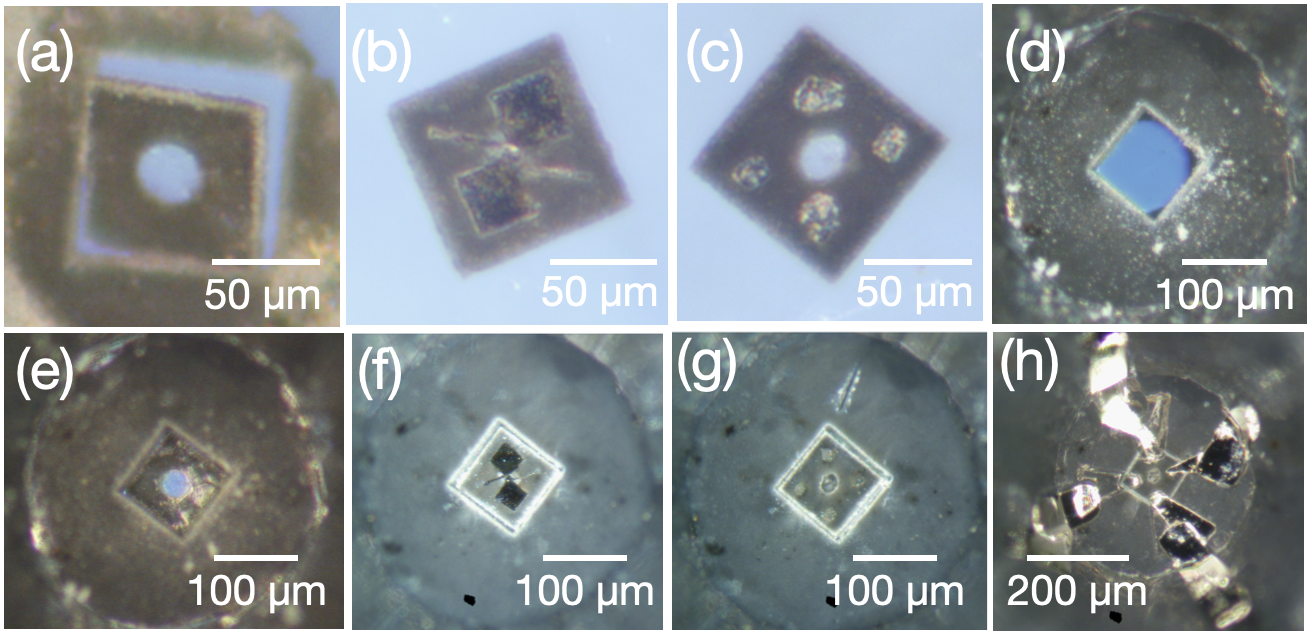}
    \caption{Loading sample \#4 into the gasket hole on a 100~$\mu$m culet. (a,b,c) The bottom, middle, and top slabs. (d) Gasket with square hole. (e) After adding the bottom slab. (f) After adding the middle slab. (g) After adding the top slab. (h) After making the final electrical connections with many pieces of platinum.}
    \label{fig:100um}
\end{figure*}

\subsection{Compression and heating}
\label{section:compression_and_heating}

Pressure is increased by tightening screws. Pressure is measured using ruby fluorescence up to~$\sim 50$~GPa, and the diamond anvil's Raman edge above~$\sim 20$~GPa.

To make final electrical connections, we typically fasten a copper-clad board to the body of the diamond cell for strain relief, and solder copper wires to and from four sections of the board (Fig. \ref{fig:closing}k-l). Our boards also have SMA connectors and barrel connectors for integration with the electrical pulser and voltage probes described in Ref. \cite{Geballe2023Pulser}. 

The samples are compressed while monitoring visually for short circuits. They are also monitored for short circuits and open circuits using a Keithley Sourcemeter 2400 to measure four-point probe resistance. Pulsed Joule heating and simultaneous measurements of four-point probe resistance and spectroradiometric temperature are performed using the methods in Refs. \cite{Geballe2023Pulser,Geballe2023PMT}. The detailed results of temperature and resistance measurements will be published elsewhere. Here, we focus on the evolution of the shape of samples and hotspots upon compression and heating. 

\section{Results}
\begin{figure*}[tbhp]
    \centering
    \includegraphics[width=3.6in]{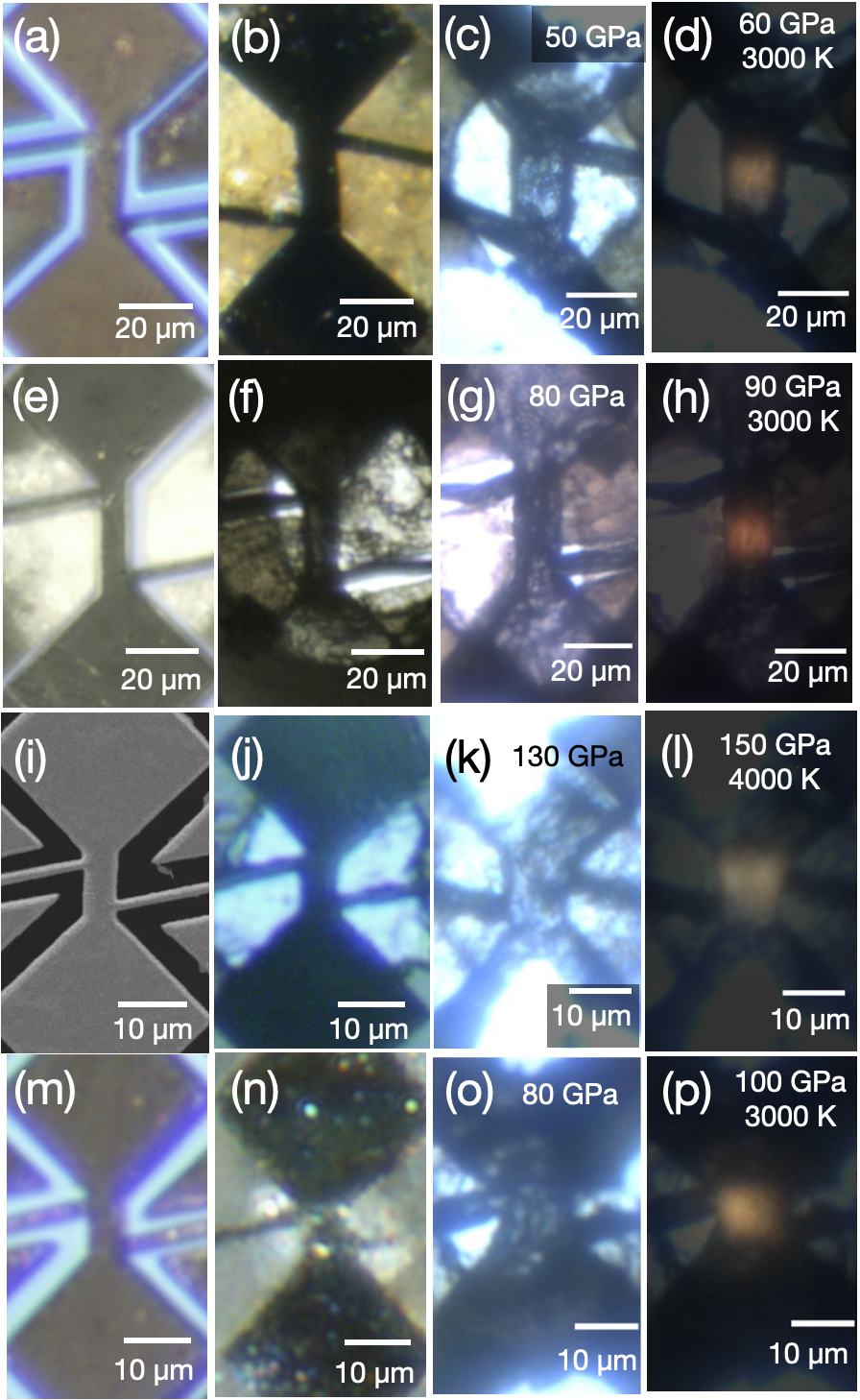}
    \caption{Four samples that successfully heat up to at least 3000 K using pulsed Joule heating at~$\sim 60$~to 150 GPa. Rows 1-4 show samples \#1-4, respectively. (Column 1) Iron samples before any compression. Panel (e) shows the iron embedded in the cBN and alumina slab; panels (a,i,m) show each iron sample surrounded by the slab of iron from which the sample was FIB-milled. (Column 2) Iron samples after a small amount of compression to the range 0-5 GPa. (Column 3) Iron samples at high pressure, illuminated in white light after pulsed Joule heating. (Column 4) Composite images made by averaging the third column images and an image of each sample's thermal emissions during pulsed Joule heating. Samples \#1-3 are successful four-point probes (i.e. not short circuited, not open circuited) up to the respective pressures and temperatures listed in the third and fourth columns. A pair of electrical leads in sample \#4 short circuited at~$40 \pm 10$~GPa, eliminating the chance to make four-point probe measurements. Panel (i) shows an SEM image; all other images are optical.}
    \label{fig:results}
\end{figure*}

The main result is simple: three samples (\#1, 2, and 3) were successfully compressed and Joule-heated to the pressure and temperature range 50-150 GPa and 2000-4000 K while measuring resistance in a four-point probe configuration (Fig. \ref{fig:results}). A fourth sample (\#4) was successfully compressed and Joule-heated to 100 GPa and 3000 K, but the four-point probe geometry was destroyed by a short-circuit upon compression. Samples \#1 and \#2 were loaded between 200~$\mu$m culets, while samples \#3 and \#4 were loaded between 100~$\mu$m culets. 

More detailed results reveal certain strengths and weaknesses of the loading procedure. Preparation of sample \#2 used two different strategies as compared to sample \#1. The loading for sample \#2 was successful, showing flexibility of the loading procedure, but the strategies were not ideal. First, the slabs for sample \#2 were cut into circles rather than squares, requiring more careful rotational alignment of the middle and top slabs. Second, the bottom slab was laser-milled into a relatively small circle, giving extra clearance compared to the gasket hole -- see Appendix \ref{section:slop}. Our motivation was to ensure easy placement of the slab into the hole. Unfortunately, the extra clearance allowed the bottom slab to slide off to one edge of the hole during the loading process, resulting in a near-overlap of cBN and the part of the iron sample that becomes hot during Joule heating (Figs. \ref{fig:results}g, l, q).

Sample \#3 was loaded between 100~$\mu$m culets with a 300~$\mu$m bevel region. All procedures followed the recipe for sample \#1, but with different dimensions (Table \ref{table:dimensions}). Upon the initial compression, one of the electrodes was not connected to the circuit. Luckily, at 60 GPa, all connections were finally complete, allowing four-point probe measurements. Sample \#4 was also loaded between 100~$\mu$m culets (Fig. \ref{fig:100um}). In an attempt to avoid open circuits, we used larger pieces of platinum for sample \#4 compared to sample \#3. Unfortunately, during compression from 30 to 50 GPa, one of the platinum pieces from the top slab appeared to short circuit with one of the iron electrodes in the middle slab, eliminating the chance to perform four-point probe resistance measurements above 30 GPa for this sample. In summary, one experiment with 100~$\mu$m culets resulted in an open circuit below 60 GPa while another resulted in a short circuit at~$40 \pm 10$~GPa, suggesting a small margin for error in our 100~$\mu$m culet recipe. A major reason for the small margin of error is that our machining tolerances seem to be marginal -- both when laser-milling and when hand-cutting with a razor blade (Appendix \ref{section:slop}). In addition, a different shape of FIBed iron might improve the success rate of making connection to the platinum electrodes. For example, the two narrow iron electrodes could flare out from their 2~$\mu$m inner connection to a much wider area (e.g. 20~$\mu$m) in the outer region.

\begin{figure*}[tbhp]
    \centering
    \includegraphics[width=4in]{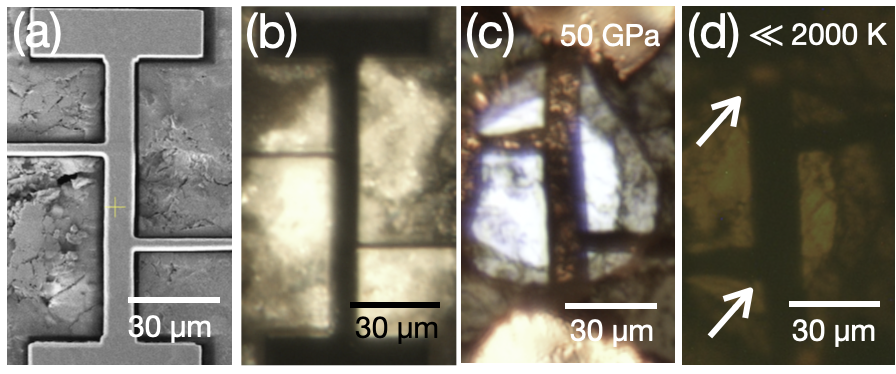}
    \caption{Sample \#5, using a different shape for the iron sample. A successful four-point configuration is maintained to 50 GPa, but the iron heats up outside the central region during pulsed Joule heating (d). Frames (a-d) correspond to columns 1-4 in Fig. \ref{fig:results}. (a) An SEM image of the sample embedded in alumina using the FIB ``lift-in'' technique. (b) An optical image of the sample after gentle compression between auxiliary anvils. (c) After compression to 50 GPa. (d) A composite image of the 50 GPa, room temperature sample in white light and thermal emissions during pulsed Joule heating. White arrows point to the very low-intensity thermal emissions.}
    \label{fig:TheLiberal}
\end{figure*}

Fig. \ref{fig:TheLiberal} shows the result of an earlier design that uses an alternative sample shape. A four-point probe configuration with relatively narrow leads and a long and narrow sample region were maintained up to 50 GPa. Unfortunately, the pulsed Joule heating hotspots at 50 GPa were located outside of the central region  (white arrows in Fig. \ref{fig:TheLiberal}d). These regions were not suitable for clean experiments because they are compressed between layers of cBN (which is mixed with epoxy). Moreover, they could not be monitored with a four-point probe.

Finally, appendix \ref{section:pure_media} shows two examples of messy, relatively unsuccessful loadings. One loading used alumina to surround the sample from all sides (i.e. no KCl). The other loading used KCl to surround the sample from all sides (i.e. no alumina). Each loading ended up with short circuits at high pressure, eliminating the chance to make four-point resistance measurements. Nonetheless, each sample was successfully pulsed Joule heated to 3000 K. In each case, the cause of the short circuit is unrelated to the choice of pressure medium. 

\section{Discussion}
The three layer recipe presented here involves many steps, each of which is simple enough to be explained by photograph. By contrast, the traditional loading method that we used in Ref. \onlinecite{Geballe2021} involves fewer steps, fewer engineering controls, and requires more subtle manipulations that are difficult to describe by photograph. For example, a key step in our previous loading method was arranging four pieces without securing any of them -- a platinum sample, a KCl insulation layer, and two platinum electrodes. Next, the four pieces were pressed while monitoring for unwanted slippage of any piece. Many iterations were necessary, involving micromanipulation of pieces that had slipped and addition of new pieces to fill gaps. It would be difficult to present a simple sequence of photographs that captured the variety of micromanipulations necessary in practice. In contrast, the new three layer recipe essentially solves this slippage problem by the fabrication of slabs. The main disadvantage of the three layer recipe is that the large number of steps requires a large amount of time. 

The results of the three layer recipe suggest unprecedented control of electrical fields and Joule-heated temperature fields in DACs. First, the wide, well-centered hotspots documented for samples \#1-4 rival the wide Joule-heated hotspot documented in Fig. 2 of Ref. \onlinecite{Zha2008}. Second, the narrow electrical leads surrounding samples \#1-3 are even narrower than the electrical leads in the breakthrough studies of Refs. \onlinecite{Ohta2016, Inoue2020, Ohta2023}. Third, Joule heating and four-point probe resistance measurements can be performed simultaneously for samples \#1-3, a capability that has not been demonstrated at pressure~$> 50$~GPa using any technology including designer diamond anvils, to the best of our knowledge. Fourth, the reproducibility of the three layer loading method is documented through photographs that show the shape of samples and of Joule-heating hotspots (Fig. \ref{fig:results}). Moreover, the hotspots are not saturated. For comparison, all photos of hotspots in Refs. \onlinecite{Ohta2023, Komabayashi2009, Ohta2016, Inoue2020, Weir2009, Weir2012, Zha2003, Zhang2020} appear to be saturated, and Refs. \onlinecite{Zhang2022,Zhang2021, Sinmyo2019,Boehler1986,Liu1975} show no photos of hotspots whatsoever. The lack of documentation in previous publications makes it very difficult to know how hotspot size compared to the region being probed -- i.e., the region between electrodes for resistance measurements, and the region probed with spectroradiometry for temperature measurements. 

The three layer assembly method is the crucial innovation in this study. When combined with FIB-embedding, and electrode holders, it allows for the reproducibility shown in Fig. \ref{fig:results}. In addition, the three layers allow for independent choices of the two thermal insulation layers and the electrical insulation that buttresses the innermost electrodes. In our implementation, the independent choice allowed us to limit sample deformation with the alumina middle layer, while simultaneously limiting heat losses with the KCl top and bottom layers.

Like the designer diamond anvils, the three layer assembly method allows for control of geometry (e.g. samples \#1-3, and to some extent, samples \#4-5). Compared to designer diamond fabrication, the three layer methods uses equipment that is much more common and somewhat less expensive -- a focused ion beam and a laser mill for the three layer method, versus lithography, sputtering, chemical vapor deposition, and etching equipment for designer diamond anvils. The three layer method allows for electrodes that are relatively thick ($\sim 5-10$~$\mu$m starting thickness), and which are made from high-purity wires, as opposed to sputtered films. The thickness and purity are helpful in allowing the 10s of A currents required for microsecond-timescale pulsed Joule heating of our samples. 

The main downside of the three layer method compared to designer anvils is large amount of human time and FIB time required to make the three slabs. The amount of time depends on a person's experience. Even the most experienced person can spend ten hours making a middle slab, plus twenty additional hours waiting for the FIB to mill parts. Bottom slabs are the simplest. They can take as little as one hour of human time per slab and do not require a FIB. Because of the substantial time investment in slab fabrication, it is crucial to have a high success rate when transferring each slab to the real gasket and stacking them atop one another.

There are many possible applications of layered microassemblies for electrical measurements beyond four-point probe measurements. For example, they could simplify existing procedures or enable new innovations for NMR, ODMR, magnetic susceptibility, and Seebeck coefficient measurements in diamond cells. \cite{Meier2017,Hsieh2019,Jackson2005, Yuan2014}

Finally, it is possible that layered microassemblies for DACs could employ more than three layers. For example, a five layer assembly made from KCl-Ir-sample-Ir-KCl could allow for uniform heating of silicates and oxides with Joule heating, a method analogous to the assemblies used for petrology in multi-anvil experiments. 

\section{Conclusions}

A reproducible recipe for a three layer microassembly has been demonstrated for the preparation of samples for Joule heating and four-point probe electrical measurements in diamond anvil cells. The method uses a bottom layer with cBN and KCl, a middle layer with a FIB-milled sample embedded in alumina and surrounded by cBN, and a top layer with four squares of platinum and a disc of KCl embedded in cBN. The layers are assembled in a stack, connected electrically to the edge of the DAC, compressed, and pulsed-Joule heated up 150 GPa and 4000 K. Successful fabrication and compression leads to many new opportunities for experiments with Joule heating and/or electrical measurements.


\begin{acknowledgments}

This material is based upon work supported by the National Science Foundation under Grant No. 2125954. We thank Amol Karandikar and Maddury Somayazulu for fruitful discussions, Matthew Diamond for helpful comments on the manuscript, and Seth Wagner, Vic Lugo, and Tyler Bartholomew for machining parts.

\end{acknowledgments}

\appendix

\section{cBN}
\label{section:cBN}
The mixture of cubic boron nitride and epoxy is made using a variation on the method of Refs. \onlinecite{Funamori2008, Wang2011}. Epotek 353ND Parts A and B are mixed in the 10:1 ratio specified by the manufacturer. A~$\sim 2$~mm drop ($\sim 50$~mg) is placed on a glass slide. A hard plastic stick with a blunt end ($\sim 2$ mm diameter) is used to mix in cBN powder (0.25~$\mu$m from Advanced Abrasives Corporation). The powder is added little by little, mixing thoroughly after each addition. The final ratio is approximately 1:10 by weight (e.g. 50 mg epoxy, 500 mg cBN), but the actual determination that enough cBN has been added is qualitative: the mixture does not seem like it can accommodate any further cBN, and the hand of the person mixing is very tired (e.g. after 20 mins of stirring). For comparison, Refs. \onlinecite{Funamori2008, Wang2011} simply reports using a 1:10 ratio of epoxy to cBN.

Next, the mixture is left to dry in air for at least 24 hours. The mixture is stored in air at room temperature. One batch can be used for several years.  

\section{Loadings with a medium of pure alumina or pure KCl}
\label{section:pure_media}
\begin{figure*}[tbhp]
    \centering
    \includegraphics[width=4in]{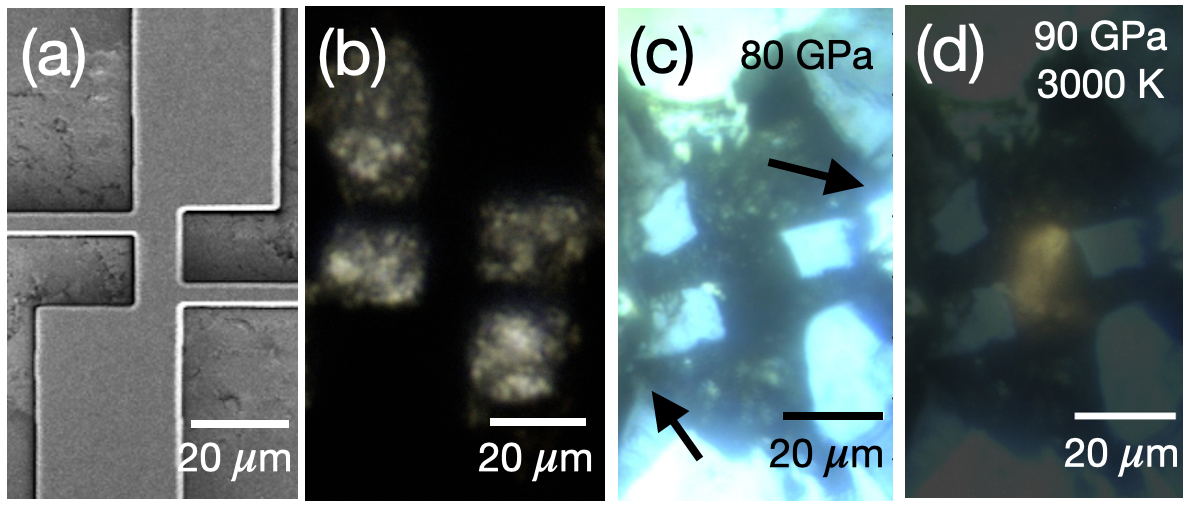}
    \caption{Sample \#6 with a 100\% alumina pressure medium, and a different shape for the iron sample. (a) An SEM image of the sample embedded in alumina using the FIB ``lift-in'' technique. (b) An optical image after loading the cell and compressing to~$\sim 5$~GPa. (c) After compression to~$\sim 80$~GPa. The black arrows point to the short circuits. (d) A composite image of (c) and and image of the thermal emissions during pulsed Joule heating to~$\sim 3000$~K.}
    \label{fig:pure_alumina}
\end{figure*}

Sample \#6 used alumina for all three layers, and a slightly different shape of the iron sample (Fig. \ref{fig:pure_alumina}a). The sample Joule heated to 1000s of K at~$\sim 70$~GPa, but short circuits eliminated the chance to make four-point probe resistance measurements. The short circuits were likely caused by the non-standard sample shape.

\begin{figure*}[tbhp]
    \centering
    \includegraphics[width=5in]{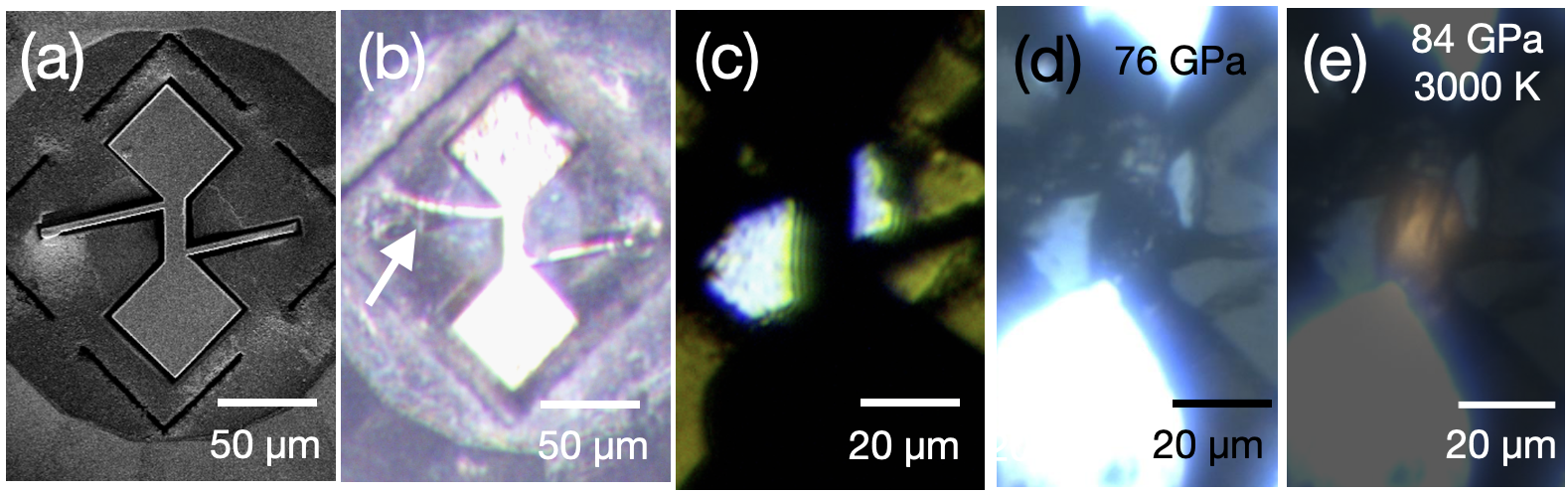}
    \caption{Sample \#7 with a 100\% KCl pressure medium and two slabs. (a) An SEM image of a $\sim 10$~$\mu$m-thick sample embedded in a ~$\sim 22$~$\mu$m-thick piece of cBN and KCl using the FIB ``lift-in'' technique. (b) An optical image after pressing the square from (a), along with a small piece of KCl (not shown), into the gasket hole. The white arrow points to the iron electrode that was bent during the pressing process. (c) After compression to~$\sim 5$~GPa. (d) After compression to 76 GPa. A short circuit formed out of view. (e) A composite image of (d) and thermal emissions during pulsed Joule heating to~$\sim 3000$~K.}
    \label{fig:pure_KCl}
\end{figure*}

Sample \#7 used a pure KCl medium. In this case, we attempted to use two layers rather than three. The idea was to integrate the bottom and middle layers - a simple idea that led to a cascade of problems, including the bent electrode in Fig. \ref{fig:pure_KCl}b. The end result was a sample that heated to 3000 K, but which contained one short circuit and one very broad inner electrode.


\section{Re-using outer electrodes}
\label{section:re-use}

Outer electrodes can be re-used many times for many high pressure runs. After a high pressure run, we typically observe thin and relatively brittle platinum arrowhead. We simply break-off the thin region using a needle, or remove it by slicing with a curved scalpel. Next, we refurbish each arrowhead using one of several strategies. The quickest option is to use a curved scalpel to reshape the arrowhead. This option only works if a flat region of suitable thickness remains on the arrowhead. A more time-consuming and more robust option is to remove the electrode holder and reshape the arrowhead with a DAC outfitted with 1 mm anvils, plus a scalpel. The most robust option is to remove the electrode holder, desolder the copper foil that holds the platinum arrowhead, and reshape the arrowhead with 1 mm anvils and a razor blade.

\section{Outer electrode recipe for other types of diamond cells}
\label{section:other_cells}
Our electrode holder is simple to fabricate and use because of the relatively large inner cavity of the piston of the Zha cell.\cite{Zha2017} BX-90 cell pistons have similarly large inner cavities. Indeed, we have successfully used a cross-shaped electrode holder that slides into our BX-90 piston. Mao-Bell cells have small but easily accessible inner cavities, which allowed us to use electrode holders in Ref. \onlinecite{Somayazulu2019}.

In contrast, symmetric cell pistons have small inner cavities. The small size allows little room for adhesives (e.g. sticky tack) to fix a loose electrode holder, and little surface area upon which a tight-fitting electrode holder can slide in and out. In addition, the portholes of a symmetric cell are small (5 mm diameter) and far from the cell's outer edge (14 mm), which makes it a major challenge to create electrical connections through the portholes after closing the cell. For comparison, Zha cell's portholes have 12 mm diameter and are located a mere 2 mm from the outer edge of the cell, making it very easy to solder electrical connections after closing the cell.

\section{Laser- and razor- machining precision}
\label{section:slop}
\begin{figure*}[tbhp]
    \centering
    \includegraphics[width=2.5in]{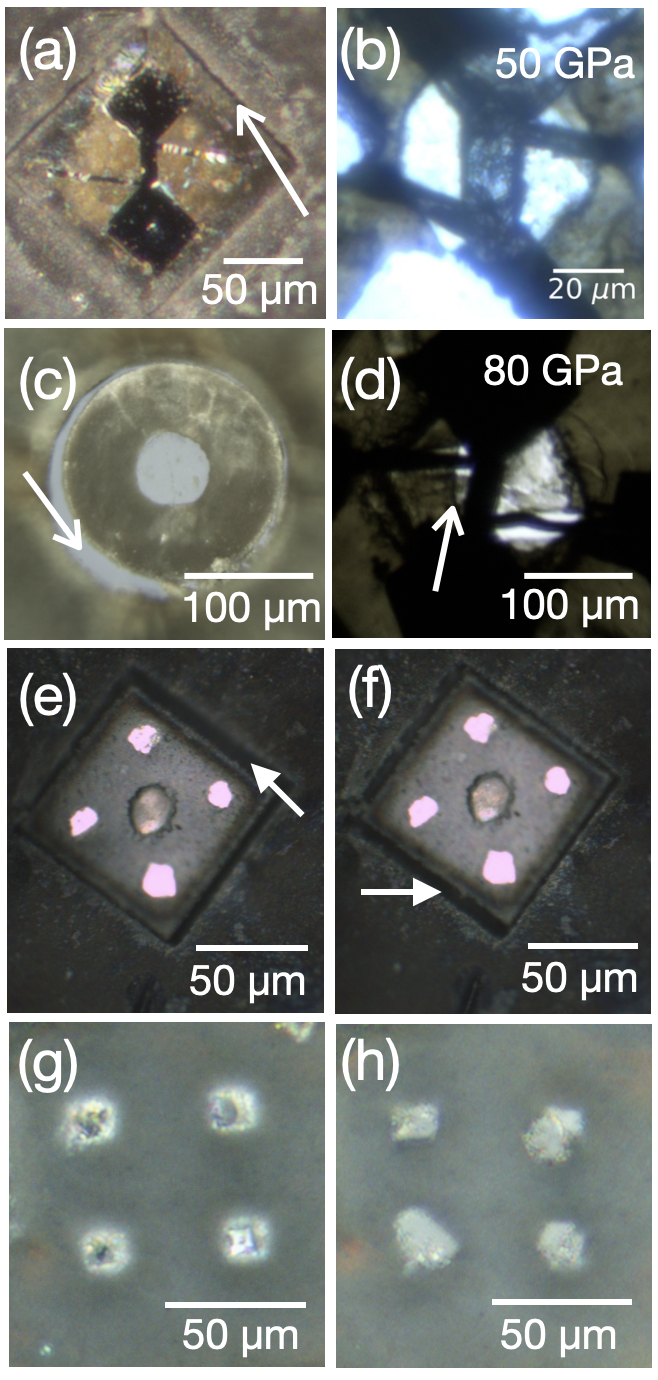}
    \caption{Four examples that provide constraints on the machining tolerances that are achieved and required. (a-b) Sample \#1. The tolerance in machining the middle slab and the gasket is approximately 6~$\mu$m, as indicated by small gap marked by the white arrow. The tolerance is tight enough to achieve well-centered samples and gasket holes at 50 GPa (b). (c-d) Sample \#2. The design choice to undersize the bottom slab by~$\sim 20$~$\mu$m (white arrow in c) results in the slab sliding off-center with respect to the middle slab and nearly touching the iron sample (white arrow in d). (e-f) Sample \#4. The tolerance in machining the top slab and the gasket is~$\sim 6$~$\mu$m; white arrows show the gap on one side in (e) and after pushing the slab slightly, the gap on the other side in (f). (g-h) Electrode fabrication for sample \#4. After a few iterations of placing platinum and pressing platinum, the underfilled holes in (g) become overfilled holes in (h). The tolerance achieved when hand-cutting, placing, and pressing pieces of platinum is apparently~$\sim 10$~$\mu$m -- see also Fig. \ref{fig:dimensions}d.}
    \label{fig:slop}
\end{figure*}

The precision of laser milling is crucial for the three-layer assemblies. If there is too little clearance for the bottom and middle slabs in the gasket hole, then force must be applied (e.g. by micromanipulator needles) to try to push the slabs into the hole. The slabs are fragile due to their aspect ratio ($\sim 10 \times 140 \times 140$~$\mu$m), so they can easily break if pressed. If there is too much clearance, on the other hand, slabs can slide with respect to one another. For example, the bottom slab for sample \#2 appears to have $\sim 10$~$\mu$m of extra clearance on each side (Fig. \ref{fig:slop}c). The improved design used for sample \#1 reduced the clearance to approximately 3~$\mu$m on each side (Fig. \ref{fig:slop}a). The 3~$\mu$m of clearance per side suggests that a slab can translate approximately 6~$\mu$m from one side to the other. Indeed, a pair of images of the top slab of sample \#4 shows that it can indeed translate 6~$\mu$m when pushed gently with a micromanipulator (Fig. \ref{fig:slop}e-f. This is apparently more than enough precision for the recipe for 200 micron culets. In contrast, the open circuit below 60 GPa for sample \#3 and the short circuit above~$\sim 40$~GPa for sample \#4 suggest that the precision is marginal for 100~$\mu$m culets.

To begin to quantitatively understand the margin for error, we first estimate the placement accuracy needed for 200 micron culets and 100 micron culets. At a minimum, the required placement accuracy is the typical distance between platinum in the top slab and the wide, rhombus-shaped iron electrodes, after compression to typical pressures, say 80 GPa and 100 GPa, respectively. The typical distances are $\sim 25$~$\mu$m, and~$\sim 12$~$\mu$m, respectively. 

\begin{figure*}[tbhp]
    \centering
    \includegraphics[width=5in]{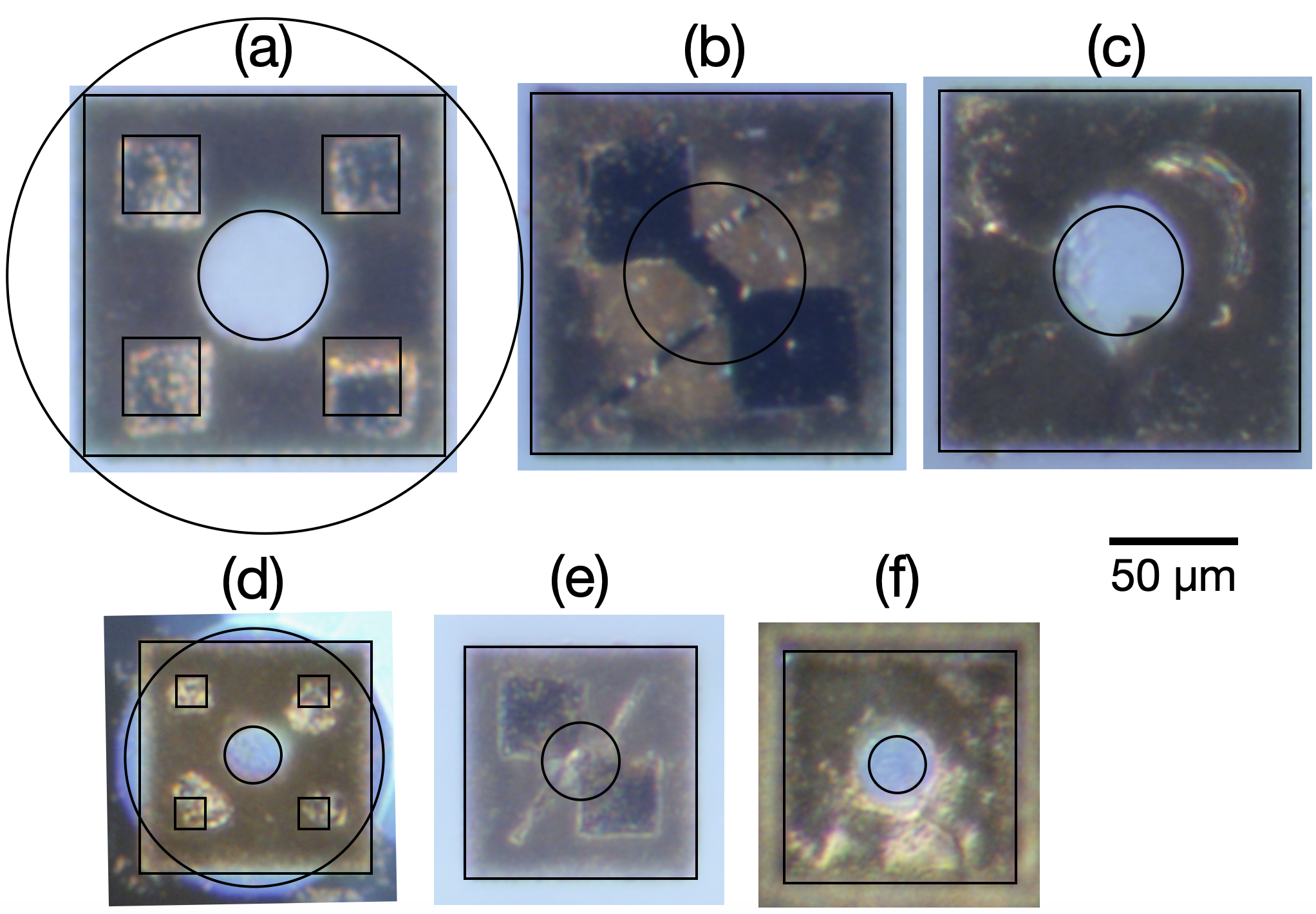}
    \caption{The ideal dimensions listed in Table \ref{table:dimensions} overlaid onto photos of the actual slabs fabricated for samples \#1 (a-c) and \#4 (d-f). Large circles in (a) and (d) are the 200 and 100~$\mu$m culet dimensions, respectively. The scale bar is the same for all panels.}
    \label{fig:dimensions}
\end{figure*}

Two types of machine tolerances - from laser milling and razor cutting - can each approximately explain the 12~$\mu$m of slop needed to justify why the 100~$\mu$m recipe works marginally. The most sensitive applications of laser machining are cutting out the top and middle slabs, and cutting the square hole in the gasket. Empirically, our recipe generates~$\sim 6$~$\mu$m of slop between each slab and the gasket (Fig. \ref{fig:slop}e-f), perhaps because this is the tightest fit that avoids slabs breaking during assembly, or perhaps because our recipe uses a bit of unnecessary clearance. Together, the middle slab-to-gasket slop and top slab-to-gasket slop could generate up to 12~$\mu$m of slab-to-slab slippage. Similarly, small imprecision in razor-cut platinum sizes can cause us to use platinum pieces that underfill the laser-cut holes (e.g. Fig. \ref{fig:slop}g), or to overfill their holes (e.g. Fig. \ref{fig:slop}h). In the end, the platinum pieces in a typical top slab are misplaced or overfilling their holes by up to $\sim 10$~$\mu$m (Fig. \ref{fig:dimensions}a,d). In this way, we can approximately explain~$\sim 12$~$\mu$m slop in spite of the fact that each type of machining is capable of a couple~$\mu$m precision on each side of a piece.

\begin{table*}[ht!]
  \centering
  \begin{tabular}{ |  m{4cm} | m{6cm} | c | }
    \hline
   Anvils	&	Two pieces. 200~$\mu$m culet, 2 mm table	&     \begin{minipage}{.3\textwidth}
      \includegraphics[width=1.2in]{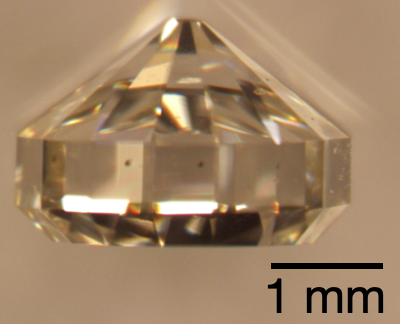}
    \end{minipage} \\ \hline 
Seats	&	Two pieces. 1 mm aperture, 60$^\circ$ full opening, 6.5 mm height, 11.5 to 13 mm diameter  (tapered)	
&     \begin{minipage}{.3\textwidth}
      \includegraphics[width=1.0in]{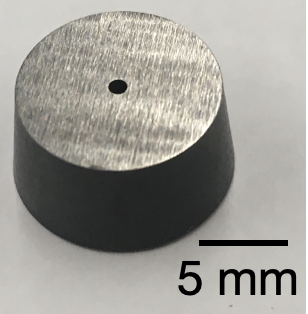}
    \end{minipage} \\ \hline
DAC	&	Zha-type	&     \begin{minipage}{.3\textwidth}
      \includegraphics[width=1.9in]{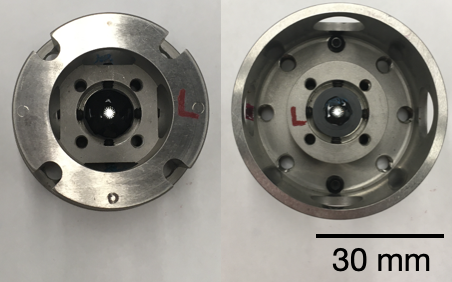}
    \end{minipage} \\ \hline
Steel disc for electrode holder	&	Machined from 1.5 mm-thick stainless steel	&     \begin{minipage}{.3\textwidth}
      \includegraphics[width=1.0in]{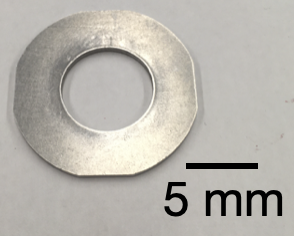}
    \end{minipage} \\ \hline
Copper-clad board	&	Four pieces of $7 \times 5$~mm~$\times 1/16$~in-thick; four pieces of~$5 \times 5$~mm~$\times 1/32$~in-thick	&     \begin{minipage}{.3\textwidth}
      \includegraphics[width=1.9in]{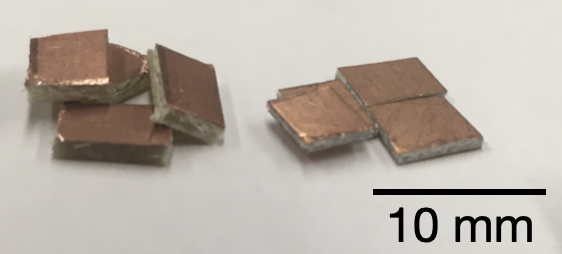}
    \end{minipage} \\ \hline
Gasket	&	Tungsten, 5 x 5 mm x 0.25 mm-thick 	&     \begin{minipage}{.3\textwidth}
      \includegraphics[width=0.8in]{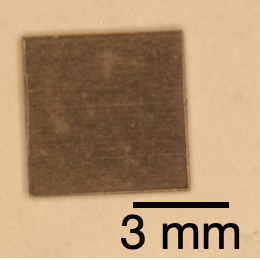}
    \end{minipage} \\ \hline
Auxiliary gaskets	&	Three pieces. 250~$\mu$m-thick, $\sim 5$mm side length & \begin{minipage}{.3\textwidth}
      \includegraphics[width=0.9in]{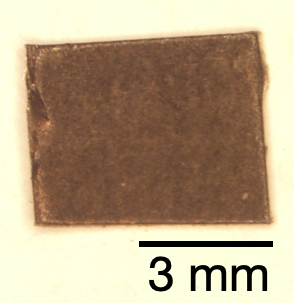}
    \end{minipage} \\ \hline
cBN	&	0.25 micron, Advanced Abrasives Corp.	&     \begin{minipage}{.3\textwidth}
      \includegraphics[width=1.0in]{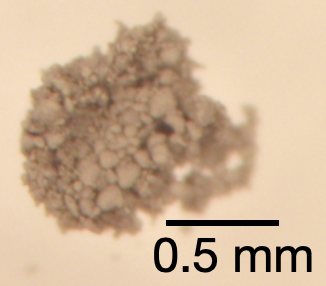}
    \end{minipage} \\ \hline
    Epoxy for anvils and copper-clad board	&	Stycast 2750FT + CAT 24LV catalyst	&     \begin{minipage}{.3\textwidth}
      \includegraphics[width=1.2in]{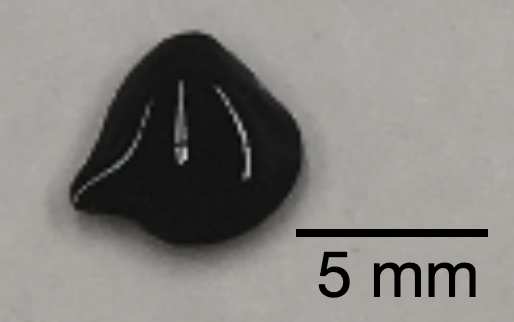}
    \end{minipage} \\ \hline
        \end{tabular}
        \end{table*}
        \begin{table*}
  \begin{tabular}{ |  m{5cm} | m{5cm} | c | }
Glue	&	Loctite Gel Control Super Glue	&     \begin{minipage}{.3\textwidth}
      \includegraphics[width=0.8in]{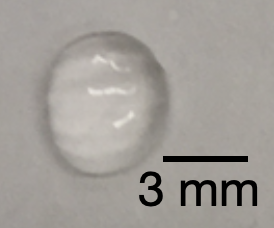}
    \end{minipage} \\ \hline
Sticky tack	&	Loctite Fun-Tack	&     \begin{minipage}{.3\textwidth}
      \includegraphics[width=0.8in]{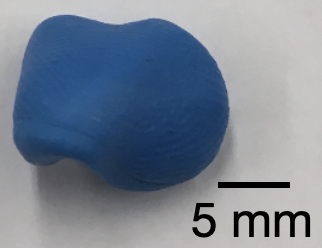}
    \end{minipage} \\ \hline
Copper wire	&	Four pieces. 20 mm long, 0.2 mm diameter. Tinned.	&     \begin{minipage}{.3\textwidth}
      \includegraphics[width=1.9in]{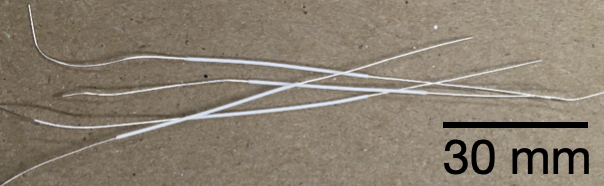}
    \end{minipage} \\ \hline
Copper foil	&	Four pieces. 9 x 2 mm x 0.2 mm-thick.	&     \begin{minipage}{.3\textwidth}
      \includegraphics[width=1.2in]{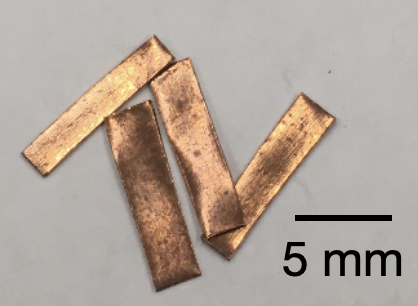}
    \end{minipage} \\ \hline
Platinum wire - thick	&	Four pieces. 0.127 mm diameter, 7 mm long.	&     \begin{minipage}{.3\textwidth}
      \includegraphics[width=1.9in]{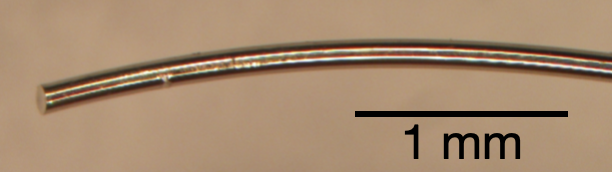}
    \end{minipage} \\ \hline
Platinum wire - thin	&	25 micron diameter. Cut into 30 to 500 micron-long pieces.	&     \begin{minipage}{.3\textwidth}
      \includegraphics[width=1.4in]{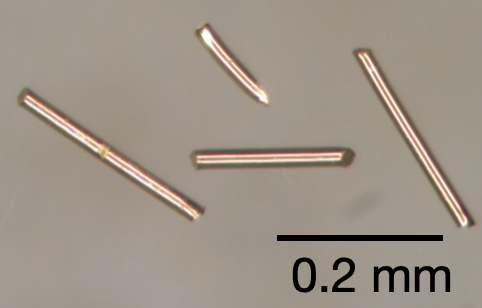}
    \end{minipage} \\ \hline
Solder	&	0.8 mm diameter with core of flux	&     \begin{minipage}{.3\textwidth}
      \includegraphics[width=1.2in]{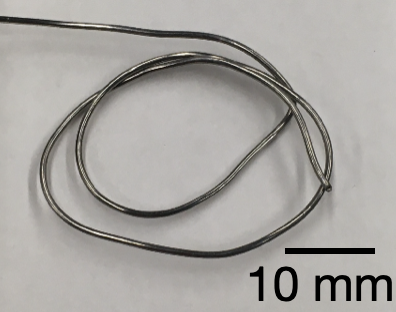}
    \end{minipage} \\ \hline
KCl	&	Sigma Aldrich, SigmaUltra,~$\geq 99.0\%$ ”	&     \begin{minipage}{.3\textwidth}
      \includegraphics[width=1.2in]{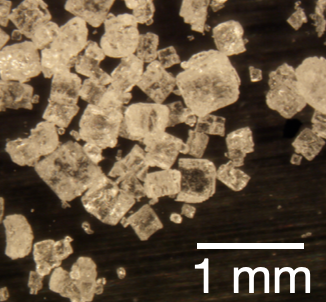}
    \end{minipage} \\ \hline
Fe	&	0.1 mm-thick foil, Alfa Aesar, Puratronic, 99.995\% metals basis &     \begin{minipage}{.3\textwidth}
      \includegraphics[width=1.2in]{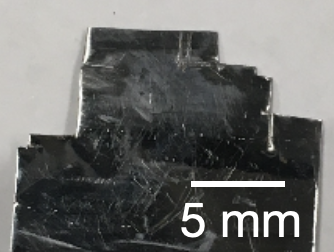}
    \end{minipage} \\ \hline

  \end{tabular}
  \caption{Ingredients.}
 \label{table:ingredients}
\end{table*}

\clearpage
\bibliography{DAC_microassembly}

\end{document}